\documentclass[11pt,noindent]{article}
\usepackage{graphics,cite,amssymb,float,ifpdf}
\usepackage{amsmath, mathbbol}

\usepackage{axodraw}
\usepackage{amsfonts}

\usepackage{epsf}
\usepackage{epsfig}

\ifpdf
\usepackage[pdftex]{graphicx}
\usepackage[pdftex,dvipsnames]{xcolor}
\usepackage[final]{pdfpages}
\else
\usepackage{graphicx}
\usepackage[dvipsnames]{xcolor}
\usepackage{psfrag}
\fi
\ifpdf
\DeclareGraphicsExtensions{.pdf, .jpg, .png}
\else
\DeclareGraphicsExtensions{.eps, .jpg}
\fi

\usepackage{array}
\usepackage{multirow}
\def\lsim{\;\raise0.3ex\hbox{$<$\kern-0.75em\raise-1.1ex\hbox{$\sim$}}\;}
\def\gsim{\;\raise0.3ex\hbox{$>$\kern-0.75em\raise-1.1ex\hbox{$\sim$}}\;}

\def\ben{\begin{enumerate}}  \def\een{\end{enumerate}}
\def\bit{\begin{itemize}}    \def\eit{\end{itemize}}
\def\beq{\begin{equation}}   \def\eeq{\end{equation}}
\def\ba{\begin{array}}       \def\ea{\end{array}}
\def\bea{\begin{eqnarray}}   \def\eea{\end{eqnarray}}

\newcolumntype{C}{ >{\centering\arraybackslash} m{4cm} }

\begin{document}

\setcounter{footnote}{0}
\vspace*{-1.5cm}
\begin{flushright}
LPT Orsay 15-71 \\
PCCF RI 15-03

\vspace*{2mm}
\today								
\end{flushright}
\begin{center}
\vspace*{5mm}

\vspace{1cm}
{\Large\bf 
Impact of sterile neutrinos on nuclear-assisted cLFV processes}
\vspace*{0.8cm}

{\bf A. Abada$^{a}$, V. De Romeri$^{b}$ and A.M. Teixeira$^{b}$} 
  
\vspace*{.5cm} 
$^{a}$Laboratoire de Physique Th\'eorique, CNRS, \\
Univ. Paris-Sud, Universit\'e Paris-Saclay, 91405 Orsay, France
\vspace*{.2cm} 

$^{b}$ Laboratoire de Physique Corpusculaire, CNRS/IN2P3 -- UMR 6533,\\ 
Campus des C\'ezeaux, 24 Av. des Landais, F-63177 Aubi\`ere Cedex, France
\end{center}

\vspace*{6mm}
\begin{abstract}
We discuss charged lepton flavour violating processes occurring in the
presence of muonic atoms, such as      
muon-electron conversion in nuclei $\text{CR}(\mu -e, \text{ N})$, 
the (Coulomb enhanced) decay of muonic atoms into a 
pair of electrons BR($\mu^- e^- \to  e^- e^-$, N), 
as well as Muonium conversion and decay, 
$\text{Mu}-\overline{\text{Mu}}$ and $\text{Mu}\to e^+ e^-$. 
Any experimental signal of these observables calls for scenarios of
physics beyond the Standard Model. In this work, 
we consider minimal extensions of the Standard Model 
via the addition of sterile fermions, 
providing the corresponding complete analytical expressions for all
the considered observables.  We first consider an ``ad hoc'' 
extension with a single sterile fermion state,
and investigate its impact on the above observables. 
Two well motivated mechanisms of neutrino mass generation are then
considered: the Inverse Seesaw embedded into the Standard Model, 
and the $\nu$MSM. 
Our study reveals that, depending on their mass range and on the
active-sterile mixing angles, sterile neutrinos 
can  give significant contributions to the above mentioned
observables, some of them even lying within present and future 
sensitivity of dedicated cLFV experiments.  
We complete the analysis by confronting our results to other
(direct and indirect) searches for sterile fermions.

\end{abstract}
\vspace*{4mm}

\section{Introduction}
\label{sec:intro}

In addition to the several theoretical caveats of the Standard Model 
(hierarchy and flavour
puzzles for instance), three observations clearly signal the need to
extend it. New Physics scenarios must be necessarily considered,
since the Standard Model (SM) cannot explain the baryon asymmetry of
the Universe (BAU), offers no viable dark matter (DM) candidate, and cannot
account for neutrino oscillation phenomena (neutrino masses and mixings).

At present, the search for New Physics (NP) is being actively carried
in many fronts: direct searches for  new states are being pursued at
high-energy colliders, and NP is also being indirectly searched for at
high-intensity facilities, looking for rare processes or deviations from
the SM predictions.  

 By construction, lepton flavour violation (LFV) is forbidden in the
SM.  By itself neutrino oscillation phenomena signal the violation
of lepton flavour in the neutral sector, and -  neutrinos being part
of the SM $SU(2)_L$ doublets which include the charged leptons
- it is only natural to expect that LFV also occurs in the charged
lepton sector (cLFV). There are numerous possible manifestations of cLFV, both
at high- and low-energies; many observables are currently
being searched for in different high-intensity dedicated facilities,
as is the case of J-PARC, PSI, SuperB factories, and many others. 

In addition to cLFV radiative ($\ell_i \to \ell_j \gamma$) and 
three-body ($\ell_i \to \ell_j \ell_k \ell_k$) decays, 
rare muon transitions can take place in the presence of nuclei: when a  
(negative) muon is stopped in matter, it can be trapped, thus forming
a ``muonic atom''. 
In this work we consider several rare LFV processes
occurring in the presence of nuclei (the most common ones being
Aluminium, Gold and Titanium), of which the most
widely investigated is the coherent $\mu-e$ conversion (see 
e.g.~\cite{Kuno:1999jp}). Numerous
experiments have already set strong bounds on the corresponding
rate~\cite{Dohmen:1993mp,Honecker:1996zf,Bertl:2006up},  
and the experimental sensitivity to CR($\mu-e$, N) is expected
to significantly improve in the near
future~\cite{Carey:2008zz,Cui:2009zz,Kuno:2013mha,Alekou:2013eta,Natori:2014yba}. 
Recently, another interesting observable has been proposed~\cite{Koike:2010xr},
which is the Coulomb-enhanced decay of a muonic atom into a pair of
electrons, $\mu^- e^- \to  e^- e^-$. The BR($\mu^- e^- \to  e^- e^-$,
N) can be strongly enhanced in the 
presence of large nuclei~\cite{Uesaka:2015qaa},
and this process will hopefully be part of the 
physics programme of the COMET experiment~\cite{Kuno:2013mha}.  
The Muonium~\cite{Pontecorvo:1957cp},  
bound state composed by a $\mu^+$ and an $e^-$, constitues a
hydrogen-like atom (despite the absence of hadronic nuclear matter). 
Formed in matter, muonium phenomena are studied in vacuum, and include
several cLFV transitions; the study of the latter has the 
advantage of being free from nuclear uncertainties. 
Here we will consider the Muonium conversion  
$\text{Mu}- \overline{\text{Mu}}$~\cite{Feinberg:1961zza} 
and its cLFV decay $\text{Mu}\to e^+ e^-$~\cite{Cvetic:2006yg}.

Many NP models can provide contributions to these observables either
by the introduction of new sources of flavour violation and/or the
presence of additional degrees of freedom (extensions of the gauge
sector and/or of the particle field content)~\cite{Raidal:2008jk}. 
In this analysis we will focus
on the study of cLFV signals arising in minimal extensions of the SM 
by sterile fermion states.  

The sterile neutrino hypothesis is strongly motivated by a number of
observations, from an interpretation of $\nu$ oscillation 
anomalies in reactor/accelerator experiments, to an
explanation of several cosmology and astrophysics issues 
(warm DM candidates, pulsar velocities, 
BAU, etc.)~\cite{Gariazzo:2015rra}.
In addition to the experimental motivation, the sterile fermion
hypothesis is very appealing from a theoretical point of view, as
these states are an integral part of numerous SM extensions aiming at
accounting for the observed neutrino masses and
mixings~\cite{Abazajian:2012ys}.  
Indeed, the most minimal extensions of the SM allowing to accommodate
neutrino data call upon the introduction of at least two
right-handed (RH) neutrinos (SM gauge singlets) providing  
a Dirac mass term for the neutral leptons, and allowing flavour
violation in leptonic charged currents. 
Should neutrinos be Majorana fermions, then a seesaw-like mechanism
emerges as a simple explanation for the smallness of their masses 
which are thus linked to a NP high-scale via natural Yukawa
couplings. Nevertheless, and despite their simplicity, high-scale seesaws
are hard to probe, since their impact on low-energy phenomenology
is negligible, and the very massive sterile states cannot be produced
at colliders. 
Low-scale seesaw realisations offer far richer experimental prospects:
provided that the couplings of the new states to the SM particles are not
excessively tiny, they can be probed both
at colliders and in high-intensity facilities.

While one can indeed consider theoretically well-motivated models, 
such as the ``standard'' type I seesaw mechanism~\cite{seesaw:I}, 
and its many variations (for example the Inverse Seesaw),
a first approach to evaluate the phenomenological impact of the 
sterile states on a given observable is to consider
an "ad-hoc" framework, independent of any model of neutrino mass 
generation. This approach simply relies
in adding a single massive Majorana sterile state to the SM, 
which would encode the effects
of a number $n_S$ of possibly existing sterile fermion states.
No hypothesis is made regarding the
mechanism of neutrino mass generation: a leptonic mixing matrix, which
must necessarily accommodate oscillation data, provides the only
connection between the interaction and the physical states.
This ``3+1'' toy model will be used in our study for a first
discussion of the phenomenological impact of the sterile fermion states 
regarding the nuclear-assisted muonic
LFV processes. 
We will also study the prospects for two well-motivated low-scale
seesaw realisations, the Inverse 
Seesaw~\cite{Schechter:1980gr,Gronau:1984ct} and the 
$\nu$-MSM~\cite{Asaka:2005an,Asaka:2005pn,Shaposhnikov:2008pf}.

For each of the above mentioned frameworks, and having taken 
into account all available experimental constraints, 
our analysis reveals that
sterile fermions can give significant contributions 
to several cLFV nuclear-assisted processes; 
depending on their mass range and on the  active-sterile mixing
angles, the new contributions can be well within present and 
future sensitivities of the corresponding cLFV experiments.  
Concerning the decay of a muonic atom ($\mu^- e^- \to  e^- e^-$),
our phenomenological analysis has shown that it has 
a strong
experimental potential 
(especially in the case of heavier targets - such as
Lead, or Uranium); should its study be pursued at
J-PARC~\cite{Kuno:2013mha,Kuno:private}, it
would offer a powerful probe into this class of 
SM extensions, especially in what concerns the Inverse Seesaw
realisation here investigated. 

Our work is organised as follows: in Section~\ref{sec:steriles} 
we  motivate and describe the three theoretical frameworks used for our
study, also summarising the different experimental and observational
constraints imposed on the sterile fermions.  
In Section~\ref{sec:cLFV:muonic} we
discuss the cLFV observables considered, presenting the analytical
formulae and the corresponding experimental status.
The results arising from the numerical studies (for each of the
considered frameworks) are collected and discussed 
in Section~\ref{sec:clfv.results}. We complete our analysis
with a thorough comparison of several observables, confronting our
results with the prospects for direct and (other) indirect searches.
Our concluding remarks are given in Section~\ref{sec:concs}.
The two appendices collect the relevant analytical expressions for
quantities (form factors, loop functions, etc.) entering the
computation of the cLFV processes under study (Appendix A), 
as well as a brief summary of the related $\mu\to e e e$ decay (Appendix B).

\section{SM extensions via sterile states}\label{sec:steriles}
Sterile fermions, such as RH neutrinos (or other fermions
which are singlets under the SM gauge group) appear as building blocks
of many SM extensions aiming at accounting for neutrino masses and their 
mixings. The existence of sterile states 
is further motivated~\cite{Gariazzo:2015rra} as they might provide an 
explanation to neutrino oscillation anomalies,
or play a r\^ole in understanding several cosmological observations
(being viable warm DM candidates, explaining pulsar
velocities, etc.). 

In addition to possibly generating Dirac and/or Majorana masses for
the light neutrinos, the sterile states can have a non-negligible
impact for a number of processes: due to the mixing with the
light (mostly active) neutrinos, the sterile fermions induce 
modifications to the SM charged and neutral currents. If
the new sterile states are not excessively heavy, and have 
sizeable mixings to the light neutrinos,  
their phenomenological imprint 
can be important - many observables will thus be sensitive to the 
active-sterile mixing couplings, and their current experimental values (or
bounds) will thus constrain such SM extensions.

\subsection{Modified leptonic interactions}\label{sec:steriles:modLag}

Let us consider an extension of the SM via $n_S$ additional 
sterile neutral (Majorana) fermions, which have non-negligible 
mixings with the active neutrinos. In this framework, 
both charged and neutral current interactions lead to the violation of
lepton flavour; the SM boson and scalar interactions with leptons are
modified as follows\footnote{See for instance~\cite{Ilakovac:1994kj} 
  for a detailed derivation of the different currents, 
  starting from explicit lepton mass matrices.}: 
\begin{align}\label{eq:lagrangian:WGHZ}
& \mathcal{L}_{W^\pm}\, =\, -\frac{g_w}{\sqrt{2}} \, W^-_\mu \,
\sum_{\alpha=1}^{3} \sum_{j=1}^{3 + n_S} {\bf U}_{\alpha j} \bar \ell_\alpha 
\gamma^\mu P_L \nu_j \, + \, \text{H.c.}\,, \nonumber \\
& \mathcal{L}_{Z^0}^{\nu}\, = \,-\frac{g_w}{2 \cos \theta_w} \, Z_\mu \,
\sum_{i,j=1}^{3 + n_S} \bar \nu_i \gamma ^\mu \left(
P_L {\bf C}_{ij} - P_R {\bf C}_{ij}^* \right) \nu_j\,, \nonumber \\
& \mathcal{L}_{Z^0}^{\ell}\, = \,-\frac{g_w}{4 \cos \theta_w} \, Z_\mu \,
\sum_{\alpha=1}^{3}  \bar \ell_\alpha \gamma ^\mu \left(
{\bf C}_{V} - {\bf C}_{A} \gamma_5 \right) \ell_\alpha\,, \nonumber \\
& \mathcal{L}_{H^0}\, = \, -\frac{g_w}{2 M_W} \, H  \,
\sum_{i,j=1}^{3 + n_S}  {\bf C}_{ij}  \bar \nu_i\left(
P_R m_i + P_L m_j \right) \nu_j + \, \text{H.c.}\ , \nonumber \\
& \mathcal{L}_{G^0}\, =\,\frac{i g_w}{2 M_W} \, G^0 \,
\sum_{i,j=1}^{3 + n_S} {\bf C}_{ij}  \bar \nu_i  
\left(P_R m_j  - P_L m_i  \right) \nu_j\,+ \, \text{H.c.},  
\nonumber  \\
& \mathcal{L}_{G^\pm}\, =\, -\frac{g_w}{\sqrt{2} M_W} \, G^- \,
\sum_{\alpha=1}^{3}\sum_{j=1}^{3 + n_S} {\bf U}_{\alpha j}   
\bar \ell_\alpha\left(
m_i P_L - m_j P_R \right) \nu_j\, + \, \text{H.c.}\,, 
\end{align}
where the different terms have been cast in the physical (mass
eigenstates) basis; $P_{L,R} = (1 \mp \gamma_5)/2$, 
$g_w$ denotes the weak coupling constant, 
$\cos^2 \theta_w = 1 - \sin^2 \theta_w = M_W^2 /M_Z^2$, and 
$m_i$ are the physical neutrino masses (light and heavy). 
The coefficients ${\bf C}_{V}$ and ${\bf C}_{A}$ 
parametrise the SM vector and axial-vector currents 
for the interaction of neutrinos with charged leptons,  
${\bf C}_{V} = \frac{1}{2} + 2 \sin^2\theta_w$ and 
${\bf C}_{A} = \frac{1}{2}$.
In Eq.~(\ref{eq:lagrangian:WGHZ}), the indices 
$\alpha = 1, \dots, 3$ denote the flavour of the charged leptons,
while $i, j = 1, \dots, 3+n_S$ correspond to the physical (massive) 
neutrino states. 

The mixing in charged current interactions is parametrised by a
rectangular 
$3 \times (3 +n_S)$ mixing matrix, ${\bf U}_{\alpha j}$ (which 
corresponds to the (unitary) PMNS matrix, $U_\text{PMNS}$, in the
case of only three neutrino generations). 
The mixing between the left-handed leptons, which we denote by
$\tilde U_\text{PMNS}$, corresponds to a $3 \times 3$ 
block of ${\bf  U}$. It proves convenient to parametrise the 
$\tilde U_\text{PMNS}$ in terms of a matrix 
$\eta$~\cite{FernandezMartinez:2007ms}, which encodes the deviation of
$\tilde U_\text{PMNS}$ from
unitarity~\cite{Schechter:1980gr,Gronau:1984ct}, 
due to the mixing between the active neutrinos and the extra fermion states: 
\begin{equation}
\label{eq:U:eta:PMNS2}
U_\text{PMNS} \, \to \, \tilde U_\text{PMNS} \, = \,(\mathbb{1} - \eta)\, 
U_\text{PMNS}\,.
\end{equation}
For the purpose of phenomenological analyses, 
the invariant quantity $\tilde \eta$, which is defined as
\begin{equation}\label{eq:def:etatilde}
\tilde \eta = 1 - |\text{Det}(\tilde U_\text{PMNS})| \, ,
\end{equation}
is often used to illustrate the effect of the new
active-sterile mixings (corresponding to a deviation from unitarity of
the $\tilde U_\text{PMNS}$).

Notice that the extra neutral Majorana fermions can also lead to a
violation of lepton flavour in neutral currents. As seen from
Eq.~(\ref{eq:lagrangian:WGHZ}), the latter 
effect is encoded in the squared mixing matrix
\begin{equation}
\label{eq:Cmatrix:def}
{\bf C}_{ij} \,=\,\sum_{\alpha=1}^{3} {\bf U}_{\alpha i}^*\,{\bf U}_{\alpha j}\,. 
\end{equation}

\subsection{Some motivated theoretical frameworks}\label{sec:steriles:th}
A first, and simple approach, to evaluate the phenomenological
impact of the extra sterile fermions is to consider a
minimal ``toy'' model, in which one adds one massive Majorana state to
the three light neutrinos of the SM. This model-independent ``ad-hoc''
construction makes no hypothesis on the underlying mechanism of mass
generation, only assuming that the physical and the interaction
neutrino bases are related via a $4 \times 4$ unitary mixing matrix  
${\bf U}_{ij}$, whose $3 \times 4$ sub-matrix ${\bf U}_{lj}$ appears
in Eq.~(\ref{eq:lagrangian:WGHZ}). 
Thus, and in addition to the three (light) active
masses and corresponding mixing angles, it is only assumed 
that the leptonic sector contains the following degrees of
freedom: the mass of the new sterile state, $m_4$, three
active-sterile mixing angles $\theta_{i4}$, two new (Dirac) CP phases
and one extra Majorana phase.
Although not directly affected by the experimental and
observational constraints which will be mentioned in
Section~\ref{sec:steriles:constraints}, 
should the sterile state be sufficiently heavy to decay into 
a $W^\pm$ boson and a charged lepton, or into a 
light (active) neutrino and either a $Z$ or a Higgs boson, its 
decay width is indirectly bounded from the
fact that the decays should comply with a perturbative
unitarity
condition~\cite{Chanowitz:1978mv,Durand:1989zs,Korner:1992an,
Bernabeu:1993up,Fajfer:1998px,Ilakovac:1999md}, 
$\frac{\Gamma_{\nu_i}}{m_{\nu_i}}\, < \, \frac{1}{2}\, (i \geq 4)$. 
Since the dominant contribution arises from the 
charged current term, the following bound on the heavy sterile masses and
their couplings to the active 
states~\cite{Chanowitz:1978mv,Durand:1989zs,Korner:1992an,
Bernabeu:1993up,Fajfer:1998px,  
Ilakovac:1999md} is obtained:
\begin{equation}\label{eq:sterile:bounds:Ciimi}
m_{\nu_i}^2\,{\bf C}_{ii} \, < 2 \, \frac{M^2_W}{\alpha_w}\, \quad
\quad (i \geq 4)\,,
\end{equation}
where $\alpha_w=g^2_w/4 \pi$, and ${\bf C}_{ii}$ is given in
Eq.~(\ref{eq:Cmatrix:def}). 

In the numerical analysis carried in Section~\ref{sec:clfv.results}, 
we will consider a normal ordering (NH) for the light
neutrino spectra\footnote{We have also considered the 
  inverted hierarchy scheme in the numerical analysis but the 
  results (and hence our conclusions) do not qualitatively change.}; 
concerning the extra sterile state, 
we scan over the following mass range
\begin{equation}\label{eq:effective:m4range}
10^{-2} \text{ GeV }\lesssim \, m_4 \, \lesssim 10^{6} \text{ GeV }\,,
\end{equation}
while randomly varying  
the active-sterile mixing angles in the
interval $[0, 2 \pi]$ (ensuring that the condition
of Eq.~(\ref{eq:sterile:bounds:Ciimi}) is respected). 
All  the CP-violating  phases are also taken into account, and
likewise randomly varied between 0 and $2 \pi$.

\bigskip
As mentioned before, sterile fermions are an integral part of several
mechanisms of neutrino mass generation, and appear in the matter field 
content of many SM extensions (type I seesaw and its
variants, Left-Right symmetric models, GUTs). 
In our study we will also illustrate the impact of sterile fermions on
cLFV rare processes in two well motivated theoretical models: the 
Inverse Seesaw (ISS) mechanism~\cite{Mohapatra:1986bd} and the 
Neutrino Minimal SM ($\nu$MSM)~\cite{Asaka:2005an}, both
characterised by a low seesaw scale (well below the GUT
scale). Both models have a rich phenomenological and
cosmological impact, and have been subject to extensive studies in
recent years. 

\medskip
The ISS mechanism~\cite{Mohapatra:1986bd} extends the SM  
via the addition of both RH and sterile neutrinos, and allows 
to account for neutrino data with (almost) natural values of 
the Yukawa couplings and for a comparatively low seesaw scale. 
Depending on its actual realisation, the ISS does allow to accommodate
the observed DM relic abundance and (potential) indirect DM
detection hints~\cite{Abada:2014zra,Abada:2014vea}, also providing a
framework where the observed BAU can be generated via
leptogenesis~\cite{Abada:2015rta}. 

We consider here a specific ISS realisation in which 
$n_R=3$ generations of RH neutrinos 
and $n_X=3$ generations of extra singlet fermions $X$ 
are added to the SM content, both 
carrying lepton number $L=+1$~\cite{Mohapatra:1986bd}.
The modified leptonic SM Lagrangian is given by
\begin{equation}
\mathcal{L}_\text{ISS} \,=\, 
\mathcal{L}_\text{SM} - Y^{\nu}_{ij}\, \bar{\nu}_{R i} \,\tilde{H}^\dagger  \,L_j 
- {M_R}_{ij} \, \bar{\nu}_{R i}\, X_j - 
\frac{1}{2} {\mu_X}_{ij} \,\bar{X}^c_i \,X_j + \, \text{H.c.}\,,
\end{equation}
where $i,j = 1,2,3$ are generation indices and $\tilde{H} = i \sigma_2
H^*$. After electroweak (EW) symmetry breaking, the neutral lepton spectrum is
composed of three (mostly active) light states, while the 
(mostly) sterile states form three nearly degenerate pseudo-Dirac
pairs. The interaction and the physical bases are related by a 
$9\times 9$ unitary mixing matrix ${\bf U}$, which diagonalises the 
$9\times 9$ full neutrino mass matrix as ${\bf U}^T \mathcal{M} {\bf U} 
= \text{diag}(m_i)$. 
In the basis where the charged lepton mass matrix is
diagonal, the leptonic mixing matrix is given by the
rectangular $3 \times 9$ sub-matrix corresponding to the first three
columns of ${\bf U}$, with the $3 \times 3$ block corresponding to the 
(non-unitary) $\tilde U_\text{PMNS}$. 

In order to derive the contributions to the studied observables, a
scan is carried over the $9\times 9$ 
neutrino mass matrix (for a detailed
discussion of the numerical studies, see~\cite{Abada:2014cca}): 
the modulus of the entries of the $M_R$ and $\mu_X$ matrices are
randomly taken to lie on the intervals
$0.5 \text{ GeV} \lesssim (M_R)_{i}  \lesssim 10^6 \text{ GeV}$ and
$0.01 \text{ eV} \lesssim
(\mu_X)_{ij}  \lesssim 1 \text{ MeV}$, with complex entries for the 
lepton number violating matrix $\mu_X$; a modified  
Casas-Ibarra parametrisation~\cite{Casas:2001sr} 
for $Y^\nu$ allows to accommodate 
neutrino oscillation data (we take complex angles for the $R$ matrix
accounting for the additional degrees of freedom, randomly 
varying their values\footnote{The choice of a larger interval for the
complex angles would  increase the range of the entries of $Y^\nu$, and thus would lead to augmented contributions to the considered cLFV
observables; however, this would also increase the other (muonic) cLFV observables which, given the present bounds, would in turn lead to the exclusion
of these regimes.} in the interval $[0, 2\pi]$; we further verify
that the Yukawa couplings are perturbative,
  i.e. $Y^\nu < 4 \pi$). We again consider a NH for the 
light neutrino spectrum. 

\medskip
The $\nu$MSM consists in a truly minimal extension of the SM via the
inclusion of three RH neutrinos, aiming at simultaneously
addressing the problems of neutrino mass generation, 
the BAU and providing a viable DM
candidate~\cite{Asaka:2005an,Asaka:2005pn,Shaposhnikov:2008pf,
Canetti:2012kh,Drewes:2015iva}.    
 
The addition of three generations of RH Majorana states $\nu_R$ to the
SM particle content allows to add the following terms to the leptonic
Lagrangian: 
\begin{equation}\label{eq:nuMSM:Lmass}
\mathcal{L}^\text{$\nu$MSM}_\text{mass} \, = \, 
-Y^\nu_{ij}\,\bar \nu_{Ri}\, \tilde H^\dagger L_j \, -\, 
\frac{1}{2}\, \bar \nu_{Ri}\, {M_{M}}_{ij}\,\nu^c_{Rj} + \text{H.c.}\,,
\end{equation}
where $i,j=1,2,3$ are generation indices, $L$ is the $SU(2)_L$
lepton doublet and $\tilde H = i \sigma_2 H^*$; $Y^\nu$ denotes the 
Yukawa couplings, while $M_{M}$ is a Majorana mass matrix (leading to
the violation of total lepton number, $\Delta L=2$).

In addition to the three light (mostly active) neutrinos - whose
masses are given by a type~I seesaw relation - 
the neutral lepton spectrum further contains three heavy states (with
masses $m_{\nu_{4-6}}$); should the latter three play a r\^ole
regarding DM, or the BAU, their spectrum and couplings to the active
states are subject to several (strong) conditions~\cite{Canetti:2012kh,Drewes:2015iva,Canetti:2014dka}. 
As previously done in~\cite{Abada:2014cca}, we parametrise the
physical $\nu$MSM degrees of freedom in terms of its six mass 
eigenvalues, encoding all physical 
mixing angles and CP violating phases (Dirac and Majorana) in an 
effective $6\times 6$ unitary mixing matrix ${\bf U}$ (which allows  
to readily implement the already well-established bounds on the 
$\nu$MSM parameter space)\footnote{In our numerical analysis, 
  we rely on the results of~\cite{Canetti:2012kh}, where the most 
  relevant constraints are translated into bounds on the $(U^2, M)$
  planes, as well as on the splitting $\delta_M$ between the two
  heaviest states (in the range from  $\sim 10^{-4}~\rm eV \ 
  to\  1~\rm keV$). The quantity $U^2$ encodes the 
  experimentally relevant combination of couplings: 
  $U^2_{4}=U^2_{e4} + U^2_{\mu4} + U^2_{\tau4}$.}. 
The angles $\theta_{lj}$, $l=1,2,3$, $j=4,5,6$ encode the
active-sterile mixings, while the mixings between the sterile states
are given by three additional angles $\theta_{45,46,56}$. The matrix
${\bf U}$ is further parametrised by 3 additional Majorana and 9
Dirac phases. As before, we will only consider a NH for the 
light neutrino spectrum.

\subsection{Constraints on sterile fermions}\label{sec:steriles:constraints}

The potentially
sizable contributions to a number of processes, induced by the mixings
of the sterile states with the active neutrinos, implies that several
observables (low-energy, collider and cosmological) can severely
constrain these models.

Firstly, one ensures that the SM extension complies with 
{\it $\nu$-oscillation data}~\cite{Tortola:2012te,Fogli:2012ua,
GonzalezGarcia:2012sz,Forero:2014bxa,nufit,Gonzalez-Garcia:2014bfa};
in all the scenarios considered in this work, and 
for both cases of NH and inverted mass hierarchy
(IH), we require compatibility with the corresponding best-fit 
intervals~\cite{Forero:2014bxa} (no constraints being imposed on the
yet undetermined value of the CP violating Dirac phase $\delta$).

In our analysis we also apply - when applicable - {\it unitarity bounds} 
on the (non-unitarity) matrix $\eta$ (cf. Eq.~(\ref{eq:U:eta:PMNS2}));
these arise from non-standard neutrino interactions with matter,
and have been derived in~\cite{Antusch:2008tz,Antusch:2014woa,Antusch:2015mia}  
by means of an effective theory approach (valid for sterile masses
above the GeV, but  
below the electroweak scale, $\Lambda_\text{EW}$).

The addition of sterile states to the SM with a sizeable active-sterile
mixing may have an impact on 
{\it electroweak precision observables} either at
tree-level (charged currents) or at higher order. We take into account
the impact of sterile neutrinos on the invisible $Z$-decay width 
(which has been 
addressed in~\cite{Akhmedov:2013hec,Basso:2013jka,
Fernandez-Martinez:2015hxa,Abada:2013aba}), 
requiring compatibility with LEP results
on $\Gamma(Z \to \nu \nu)$~\cite{Agashe:2014kda}; 
in addition, we further require that potential 
new contributions to the cLFV
$Z$ decay width do not exceed the present uncertainty on the 
total $Z$ width~\cite{Agashe:2014kda}:
$\Gamma (Z \to \ell_1^\mp \ell_2^\pm) < \delta \Gamma_{\rm  tot}$.  

LHC data on {\it invisible Higgs} decays already allows to constrain
regimes where the sterile states are below the Higgs mass; in our
study we apply the constraints derived 
in~\cite{BhupalDev:2012zg,Cely:2012bz,Bandyopadhyay:2012px}. 
Negative {\it laboratory searches} for monochromatic lines in the
spectrum of muons from  $\pi^\pm \to \mu^\pm \nu$
decays~\cite{Kusenko:2009up,Atre:2009rg} also impose robust bounds on 
sterile neutrino masses in the MeV-GeV range. 

The introduction of singlet neutrinos with Majorana masses allows for 
new processes like lepton number violating interactions, among which 
{\it neutrinoless double beta decay} remains the most important
one~\cite{Benes:2005hn}. 
In our analysis, we evaluate the contributions of the sterile states
to the effective mass $m_{ee}$ according to~\cite{Blennow:2010th,Abada:2014nwa};
strong bounds on the effective mass have been set by several
experiments, among them GERDA~\cite{Agostini:2013mzu}, 
EXO-200~\cite{Auger:2012ar,Albert:2014awa},
KamLAND-ZEN~\cite{Gando:2012zm}. In our analysis we use 
the most recent constraint from~\cite{Albert:2014awa}.

Further constraints arise from {\it leptonic and semileptonic decays of
pseudoscalar mesons}
$\ K,\ D$, $\ D_s$, $B$  (see~\cite{Goudzovski:2011tc,Lazzeroni:2012cx} for
kaon decays,~\cite{Naik:2009tk,Li:2011nij} for $D$ and $D_S$ decay
rates, and~\cite{Aubert:2007xj,Adachi:2012mm} for $B$-meson observations).
Recent studies suggest that in the framework of the SM extended by
sterile neutrinos the 
most severe bounds arise from the violation of lepton universality
in leptonic kaon decays (parametrised by the observable 
$\Delta r_K$)~\cite{Abada:2012mc,Abada:2013aba}.

Other than the rare decays occurring in the presence of nuclei, 
the new states can contribute to several 
{\it charged lepton flavour violating processes} such as  $\ell_i \to
\ell_j \gamma$ and $\ell_i \to \ell_j \ell_k \ell_k$.
In our analysis we compute the contribution of the sterile states to
all these
observables~\cite{Ma:1979px,Gronau:1984ct,Ilakovac:1994kj,
Deppisch:2004fa,Deppisch:2005zm,Dinh:2012bp,Alonso:2012ji,Abada:2014kba},  
imposing compatibility with the bounds summarised in
Table~\ref{table:cLFV:bounds}, 
also considering the impact of the future experimental sensitivities.
\begin{table}[h!]
\centering
\begin{tabular}{|c|c|c|}
\hline
cLFV Process & Present Bound & Future Sensitivity  \\
\hline
    $\mu \rightarrow  e \gamma$ & $5.7\times
10^{-13}$~\cite{Adam:2013mnn}  & $6\times
10^{-14}$~\cite{Baldini:2013ke} \\ 
    $\tau \to e \gamma$ & $3.3 \times 10^{-8}$~\cite{Aubert:2009ag}& $
\sim3\times10^{-9}$~\cite{Aushev:2010bq}\\ 
    $\tau \to \mu \gamma$ & $4.4 \times 10^{-8}$~\cite{Aubert:2009ag}&
$ \sim3\times10^{-9}$~\cite{Aushev:2010bq} \\ 
    $\mu \rightarrow e e e$ &  $1.0 \times
10^{-12}$~\cite{Bellgardt:1987du} &
$\sim10^{-16}$~\cite{Blondel:2013ia} \\ 
    $\tau \rightarrow \mu \mu \mu$ &
$2.1\times10^{-8}$~\cite{Hayasaka:2010np} & $\sim
10^{-9}$~\cite{Aushev:2010bq} \\ 
    $\tau \rightarrow e e e$ &
$2.7\times10^{-8}$~\cite{Hayasaka:2010np} &  $\sim
10^{-9}$~\cite{Aushev:2010bq} \\ 
\hline
\end{tabular}
\caption{Current experimental bounds and future sensitivities for
  radiative and 3-body cLFV decays considered in our study.}
\label{table:cLFV:bounds}
\end{table}

Finally, a number of
{\it cosmological observations}~\cite{Smirnov:2006bu,Kusenko:2009up}
put severe constraints on sterile neutrinos with a mass below the GeV. 
While very light
sterile neutrinos (with a mass $\lesssim$ eV) appear to be disfavoured
by CMB analysis with the Planck satellite~\cite{Ade:2013zuv}, 
a sterile state with mass $\sim$keV could be a viable DM candidate, also
offering a possible
explanation for  the observed X-ray line in galaxy clusters spectra at an
energy $\sim 3.5$~keV~\cite{Bulbul:2014sua,Boyarsky:2014jta}, for 
the origin of pulsar kicks, or even to the BAU (for a review
see~\cite{Abazajian:2012ys}).  

\bigskip
Independently of the model under consideration, and
for all regimes of sterile masses investigated, 
in our phenomenological analysis we ensure that all the above referred 
constraints - theoretical (such as perturbativity of the active-sterile
  couplings) and experimental - are verified.
Since the cosmological bounds are in general derived by
assuming the minimal possible abundance (in agreement with neutrino
oscillations) of sterile neutrinos in halos consistent with standard
cosmology, 
a non-standard cosmology with a very low reheating temperature or a
scenario where the sterile 
neutrinos couple to a dark sector~\cite{Dasgupta:2013zpn}, could
allow to evade some bounds~\cite{Gelmini:2008fq}.
In our numerical analysis, we adopt a conservative approach, 
allowing for the violation
of these cosmological bounds in some scenarios (which will be
identified).

\section{Muonic atoms and cLFV rare processes}\label{sec:cLFV:muonic} 
In what follows, we present the different cLFV observables which can
be studied in relation to rare processes involving muonic atoms. 

\subsection{Muon-electron conversion}\label{sec:cLFV:muonic:CR} 
One of the most sensitive probes of cLFV is the $\mu - e$
conversion occurring in a muonic atom, which is formed when a muon is
``stopped'', falling into the 1s state of a target nucleus.  
The observable is defined as 
\begin{equation}\label{eq:CR:def}
\text{CR}(\mu -e, \text{ N}) \, = \frac{\Gamma (\mu^- + N \to e^- 
+N)}{\Gamma (\mu^- + N \to \text{ all captures)}}\,.
\end{equation}

\noindent
The coherent conversion (in which the nuclear final state is in its
ground state) increases\footnote{The behaviour is not strictly
  monotonic with $Z$; however, and although dependent of the 
  underlying source of LFV, similar patterns have been found for the
  different types of NP contribution (dipole, scalar, or 
  vector)~\cite{Kitano:2002mt}.} 
with the atomic number ($Z$) for light nuclei, 
$Z\lesssim 30$, and is maximal for $30 \lesssim Z \lesssim
60$~\cite{Kitano:2002mt}. 
For heavier nuclei, Coulomb distortion effects of the wave
function lead to a reduction of the corresponding conversion rate.  

Past and present experiments have mostly explored Titanium, Lead and
Gold nuclei; the present bounds for different targets 
(obtained by the SINDRUM II experiment) are summarised in 
Table~\ref{table:mue:bounds.past}.
\renewcommand{\arraystretch}{1.3}
\begin{table}[h!]
\centering
\begin{tabular}{|c|c|c|}
\hline
Present CR($\mu -e$, N) bound & material & year  \\
\hline
$4.3\times 10^{-12}$~\cite{Dohmen:1993mp} & 
Ti & 1993 \\
$4.6\times 10^{-11}$~\cite{Honecker:1996zf} &
Pb & 1996 \\
$7\times 10^{-13}$~\cite{Bertl:2006up} &
Au & 2006 \\ 
\hline
\end{tabular}
\caption{Current experimental bounds for CR($\mu -e$, N).}
\label{table:mue:bounds.past}
\end{table}
\renewcommand{\arraystretch}{1.}

At present, two projects are aiming at  improving
the current bounds, Mu2e (at Fermilab)~\cite{Carey:2008zz} and COMET
(J-PARC)~\cite{Cui:2009zz,Kuno:2013mha},  
sharing several common features (nominal muon beam energy 
$\sim$ 8 GeV, Aluminium targets, ...). 
The COMET experiment will carry a two-phase search for $\mu -e$
conversion, with the second phase expected to bring the sensitivity
down by two orders of magnitude with respect to Phase I. 
(It is also possible that
different targets will be used at this stage~\cite{Kuno:private}.)
The COMET experiment will be subsequently modified: PRISM/PRIME will 
further improve the sensitivity to these cLFV processes~\cite{Alekou:2013eta}.
Another possibility is that of DeeMe~\cite{Natori:2014yba}, 
which is a parallel project at J-PARC, albeit on a smaller
experimental scale: 
working with a silicon-carbide target, they aim
at lowering the present sensitivity by almost two orders of magnitude. 
In the long term, Project-X (at
Fermilab)~\cite{Nagaitsev:2014aha,Asner:2013wpa}  
is expected to benefit from
very intense muon beams, with at least ten times more muons than
Mu2e (however for a lower energy beam, around 1-3
GeV). We summarise the expected 
future sensitivities in Table~\ref{table:mue:future}.
\renewcommand{\arraystretch}{1.3}
\begin{table}[h!]
\centering
\begin{tabular}{|l|c|c|}
\hline
Experiment (material) & future sensitivity & year  \\
\hline
Mu2e (Al) & $3 \times 10^{-17}$~\cite{Carey:2008zz} & 
$\sim$ 2021 \\ 
COMET (Al) - Phase I & $3 \times 10^{-15}$~\cite{Kuno:2013mha} & 
$\sim$ 2018 \\
COMET (Al) - Phase II & $3 \times 10^{-17}$~\cite{Kuno:2013mha} & 
$\sim$ 2021 \\
PRISM/PRIME (Ti)& $10^{-18}$~\cite{Alekou:2013eta} & \\
DeeMe (SiC) & $2 \times 10^{-14}$~\cite{Natori:2014yba} & \\
\hline
\end{tabular}
\caption{Future sensitivities for CR($\mu -e$, N).}
\label{table:mue:future}
\end{table}
\renewcommand{\arraystretch}{1.}

\begin{figure}
\begin{center}
\begin{tabular}{c}
 \hspace*{-10mm}
\scalebox{0.7}{
 \begin{picture}(451,124) (71,-58)
    \SetWidth{1.5}
    \SetColor{Black}
    \ArrowLine(72.235,-43.496)(165.442,-43.496)
    \ArrowLine(72.235,39.613)(126.606,39.613)
    \ArrowLine(210.492,38.836)(258.648,38.836)
    \ArrowLine(165.442,-43.496)(258.648,-43.496)
    \Text(89.323,49.71)[lb]{\Large{\Black{$\mu$}}}
    \Text(104.857,-62.138)[lb]{\Large{\Black{$u,d$}}}
    \Text(224.473,-62.138)[lb]{\Large{\Black{$u,d$}}}
    \Text(159.228,49.71)[lb]{\Large{\Black{$\nu_i$}}}
    \Text(243.891,49.71)[lb]{\Large{\Black{$e$}}}
    \SetWidth{1.0}
    \Vertex(125.829,41.166){2.197}
    \Vertex(210.492,38.836){2.197}
    \Vertex(210.492,38.836){2.197}
    \SetWidth{1.5}
    \ArrowLine(127.382,39.613)(208.162,38.836)
    \Text(205.055,4.66)[lb]{\Large{\Black{$W^-$}}}
    \Text(110.732,4.66)[lb]{\Large{\Black{$W^-$}}}
    \PhotonArc(167.246,51.865)(44.903,-163.132,-18.951){3.884}{8.5}
    \Photon(169.325,6.214)(170.102,-43.496){3.884}{3}
    \Text(180.976,-22.525)[lb]{\Large{\Black{$\gamma$}}}
    \ArrowLine(336.321,39.613)(390.691,39.613)
    \ArrowLine(390.691,39.613)(471.47,38.836)
    \ArrowLine(472.247,38.836)(520.404,38.836)
    \ArrowLine(428.751,-43.496)(521.957,-43.496)
    \ArrowLine(334.767,-43.496)(427.974,-43.496)
    \Photon(429.527,6.99)(430.304,-42.72){3.884}{3}
    \PhotonArc(428.225,53.419)(44.903,-163.132,-18.951){3.884}{8.5}
    \Text(351.855,49.71)[lb]{\Large{\Black{$\mu$}}}
    \Text(424.09,49.71)[lb]{\Large{\Black{$\nu_i$}}}
    \Text(493.995,49.71)[lb]{\Large{\Black{$e$}}}
    \Text(370.496,-62.138)[lb]{\Large{\Black{$u,d$}}}
    \Text(479.238,-62.138)[lb]{\Large{\Black{$u,d$}}}
    \Text(372.71,7.767)[lb]{\Large{\Black{$W^-$}}}
    \Text(464.48,7.767)[lb]{\Large{\Black{$W^-$}}}
    \Text(443.508,-20.195)[lb]{\Large{\Black{$Z$}}}
  \end{picture}}
  \hspace*{13mm}
\scalebox{0.55}{ 
 \begin{picture}(252,163) (92,-73)
    \SetWidth{1.5}
    \SetColor{Black}
    \ArrowLine(93,-53)(213,-53)
    \ArrowLine(93,55)(159,55)
    \ArrowLine(271,55)(333,55)
    \ArrowLine(213,-53)(333,-53)
    \Text(113,69)[lb]{\Large{\Black{$\mu$}}}
    \Text(129,-76)[lb]{\Large{\Black{$u,d$}}}
    \Text(303,-78)[lb]{\Large{\Black{$u,d$}}}
    \Text(212,68)[lb]{\Large{\Black{$W^-$}}}
    \Text(309,66)[lb]{\Large{\Black{$e$}}}
    \SetWidth{1.0}
    \Vertex(162,56){2.828}
    \Vertex(271,53){2.828}
    \Vertex(271,55){2.828}
    \SetWidth{1.5}
    \ArrowLine(164,55)(217,11)
    \Text(259,21)[lb]{\Large{\Black{$\nu_j$}}}
    \Text(168,21)[lb]{\Large{\Black{$\nu_i$}}}
    \Photon(217,12)(217,-53){5}{3}
    \Text(233,-26)[lb]{\Large{\Black{$Z$}}}
    \SetWidth{1.6}
    \ArrowLine(217,11)(271,55)
    \SetWidth{1.5}
    \Photon(164,55)(271,55){5}{6}
  \end{picture}}  \vspace*{5mm}\\
  \scalebox{0.79}{
  \begin{picture}(451,101) (31,-40)
    \SetWidth{1.5}
    \SetColor{Black}
    \ArrowLine(31.938,-26.615)(95.815,-26.615)
    \ArrowLine(32.53,36.67)(96.406,36.67)
    \Photon(96.406,36.078)(159.691,-26.024){2.957}{7}
    \Photon(96.406,-26.615)(159.1,35.487){2.957}{7}
    \ArrowLine(160.283,36.078)(224.159,36.078)
    \ArrowLine(159.691,-27.207)(224.751,-27.207)
    \Text(41.993,44.95)[lb]{\Large{\Black{$\mu$}}}
    \Text(44.95,-41.401)[lb]{\Large{\Black{$u$}}}
    \Text(193.404,-41.993)[lb]{\Large{\Black{$u$}}}
    \Text(121.838,-44.584)[lb]{\Large{\Black{$d_j$}}}
    \Text(116.515,44.95)[lb]{\Large{\Black{$\nu_i$}}}
    \Text(191.629,47.907)[lb]{\Large{\Black{$e$}}}
    \ArrowLine(288.036,36.67)(351.912,36.67)
    \ArrowLine(288.627,-26.615)(352.504,-26.615)
    \Photon(352.504,37.261)(353.095,-26.615){2.957}{5}
    \ArrowLine(354.278,36.67)(418.154,36.67)
    \ArrowLine(353.095,-26.615)(418.154,-26.615)
    \Photon(416.972,36.67)(417.563,-27.207){2.957}{5}
    \ArrowLine(418.154,36.67)(482.031,36.67)
    \ArrowLine(417.563,-26.615)(481.439,-26.615)
    \Text(303.413,43.767)[lb]{\Large{\Black{$\mu$}}}
    \Text(309.328,-41.401)[lb]{\Large{\Black{$u$}}}
    \Text(373.204,43.767)[lb]{\Large{\Black{$\nu_i$}}}
    \Text(386.808,-43.767)[lb]{\Large{\Black{$d_j$}}}
    \Text(447.727,-42.584)[lb]{\Large{\Black{$u$}}}
    \Text(454.824,43.767)[lb]{\Large{\Black{$e$}}}
    \SetWidth{1.0}
    \Vertex(95.815,37.853){1.673}
    \Vertex(96.998,-26.615){1.673}
    \Vertex(159.1,-27.207){1.673}
    \Vertex(160.283,36.078){1.673}
    \Vertex(160.283,36.078){1.673}
    \Vertex(351.912,-27.207){1.673}
    \Vertex(353.686,36.67){1.673}
    \Vertex(419.337,36.67){1.673}
    \Vertex(416.972,-26.615){1.673}
    \SetWidth{1.5}
    \ArrowLine(96.998,36.67)(158.508,36.078)
    \ArrowLine(96.998,-26.615)(158.508,-27.207)
    \Text(152.045,10.646)[lb]{\Large{\Black{$W$}}}
    \Text(90.998,10.055)[lb]{\Large{\Black{$W$}}}
    \Text(331.169,7.689)[lb]{\Large{\Black{$W$}}}
    \Text(425.252,5.914)[lb]{\Large{\Black{$W$}}}
  \end{picture}}
  \end{tabular}
\end{center}
\caption{Box and penguin diagrams contributing to the nuclear
   $\mu-e$ conversion in the presence of (sterile) 
  massive neutrinos. In the quark internal lines, $j=1...3$ 
  runs over the three up- and down-quark families; 
  in the neutral fermion ones, $i,j=1...3+n_S$.}  
\label{Fig:muediagrams}
\end{figure}
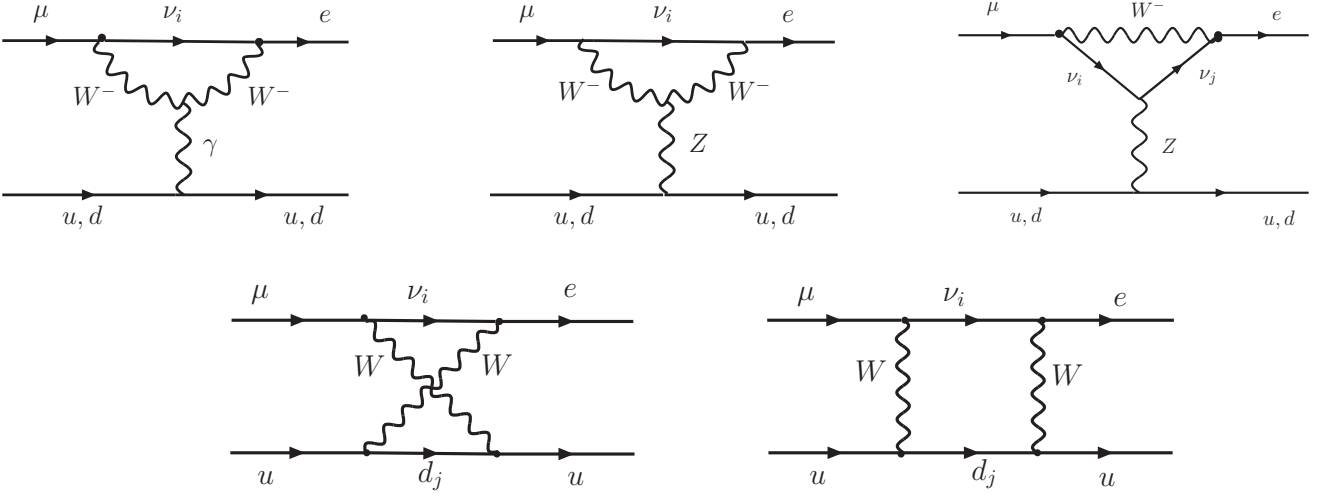

Several models of NP can give rise to the rare cLFV nuclear
conversion, and the corresponding rates have been extensively
discussed in the literature~\cite{Kuno:1999jp,Raidal:2008jk,deGouvea:2013zba}.
For the present class of SM extensions, contributions to the nuclear 
$\mu-e$ conversion arise from the $Z$- and photon-penguin diagrams, as
well as boxes, which are schematically depicted in
Fig.~\ref{Fig:muediagrams}. 
At lowest order, the flavour violating $\mu-e$ transition
originates from one-loop diagrams involving neutrinos (active and
sterile); their non-zero masses and mixings prevent a GIM cancellation
(which would otherwise occur).

The interactions depicted in Fig.~\ref{Fig:muediagrams} can be
described by the following effective Lagrangian~\cite{Alonso:2012ji}
(neglecting the electron mass), 
\begin{equation}
\mathcal{L}_\text{eff}^{\mu - e}\, =\, 
\frac{g_w^2}{2\,(4\pi)^2\,M_W^2} 
\left( \frac{\sqrt{4 \pi \alpha} }{2}\,m_\mu\, G^{\mu e}_\gamma \,\bar
e \,\sigma_{\lambda \rho} \,  \mu_R \,F^{\lambda\rho}\, 
+\, g_w^2 \,\sum_{q=u,d}  \tilde{F}_q^{\mu e}\,
\bar e \,\gamma_\rho \,\mu_L\, \bar q \,\gamma^\rho \, q \right) \,+\,
\text{H.c.}\,,  
\label{eq:efflag}
\end{equation}
where $G_\gamma^{\mu e}$ refers to the photon-lepton dipole 
coupling corresponding to an on-shell photon ("non local", long-range
contribution),   
$F^{\lambda\rho}=\partial^\lambda A^\rho-\partial^\rho A^\lambda$ 
denotes the electromagnetic field strength, and $\tilde{F}_q^{\mu e}$ 
contains the "local" contribution of the monopole $F_\gamma^{\mu e}$, 
as well as those arising from the weak gauge-boson 
exchange diagrams. 
The dependence of the process on the new degrees of freedom associated
with the sterile neutrinos are contained in the form factors 
$\tilde F_q^{\mu e}$ and in the dipole term $G^{\mu e}_\gamma$.

In the SM extended by sterile neutrinos, 
the ratio for the nuclear assisted $\mu-e$ conversion, see
Eq.~(\ref{eq:CR:def}), can be cast in the following compact form,  
\begin{equation}
\text{CR}(\mu-e, \text{N}) =
\frac{2\,G_F^2\,\alpha_{w}^2\,m_\mu^5}{(4\pi)^2
  \,\Gamma_\text{capt}(Z)}
\left|4\,V^{(p)}\left(2 \,\tilde{F}_{u}^{\mu e}+\tilde{F}_{d}^{\mu
  e}\right)+
4\,V^{(n)}\left(\tilde{F}_u^{\mu e}+2\,\tilde{F}_{d}^{\mu e}\right)
+ D \,G^{\mu e}_{\gamma} \frac{s^2_w}{2 \sqrt{4 \pi \alpha}}  \right|^2\,.
\label{eq:CRmue}
\end{equation}
In the above, $\Gamma_\text{capt}(Z)$ denotes the capture rate of the
nucleus characterised by the atomic number $Z$~\cite{Kitano:2002mt}; 
$G_F$ is the Fermi
constant, $m_\mu$ the muon mass, 
$\alpha=e^2/(4\pi)$, with $s_w$ corresponding to the sine of the weak
mixing angle. 
The  form factors $\tilde{F}_{q}^{\mu e}$ ($q=u,d$) are given by 
\begin{equation}
\tilde F_q^{\mu e}\, =\, 
Q_q \, s_w^2 F^{\mu e}_\gamma+F^{\mu e}_Z
\left(\frac{{I}^3_q}{2}-Q_q\, s_w^2\right)+
\frac{1}{4}F^{\mu eqq}_\text{Box}\,,
\label{eq:tildeFqmue}
\end{equation}
where $Q_q$ denotes the quark electric charge ($Q_u=2/3, Q_d=-1/3$) and
${I}^3_q$ is the weak isospin ($\mathcal{I}^3_u=1/2\,,\,
\mathcal{I}^3_d=-1/2$). 
The quantities $F^{\mu e}_\gamma$, $F^{\mu e}_Z$ and $F^{\mu
  eqq}_\text{Box}$ correspond to the different form factors of 
the diagrams depicted in Fig.~\ref{Fig:muediagrams}, respectively
photon-penguins, $Z$-penguins and box diagrams; their expressions are
collected in Appendix~\ref{app:A}, as is the one concerning the dipole 
term, $G^{\mu e}_\gamma$.
The relevant nuclear information (nuclear form
factors and averages over the atomic electric field) are encoded in
the  $D$, $V^{(p)}$ and $V^{(n)}$ form factors. In our analysis we use 
the numerical values presented in~\cite{Kitano:2002mt}.\\

We conclude this discussion commenting about another 
interesting nuclear-assisted LFV process, which is a 
charge-changing reaction of the type~\cite{Kuno:1999jp,Babu:1995vh}:
\begin{equation}
\mu^- + (A,Z) \rightarrow e^+ + (A,Z-2)^* \, ,
\end{equation}
denoted $\mu^- \rightarrow e^+$ conversion in nuclei. 
The final nucleus, which is different from the initial one - thus
preventing a coherent enhancement - can be either in its ground state
or in an excited one.  
This rare process violates the conservation of the total lepton number
by two units and is related to neutrinoless double decay
process. Although the sterile neutrinos may give a non negligible
contribution to the $\mu^- \rightarrow e^+$ conversion rate, we do not
address this observable here.


\subsection{Decay of muonic atoms to $\pmb{e^- e^-}$ pairs: $\pmb{\mu^- e^- \to
  e^- e^-}$}\label{sec:cLFV:muonic:mue2e} 

A different cLFV process was recently proposed in~\cite{Koike:2010xr}. 
It consists in the flavour violating decay of a bound $\mu^-$ in a
muonic atom into a pair of electrons, 
and has been identified as potentially complementary to other cLFV
muon decays:
\begin{equation}\label{eq:me2ee}
  \mu^{-} \, {e^{-}} \to\, {e^{-}}\,{e^{-}}\,.
\end{equation}
In the above transition, 
the initial states are a $\mu^-$ and a 1s atomic $e^-$,
bound in the Coulomb field of a nucleus~\cite{Koike:2010xr}.  

Although the underlying sources of flavour violation giving rise to
this observable are the same as those responsible for other
non-radiative $\mu-e$ transitions (such as $\mu \to e e e$, or Muonium
decay), the $\mu^- e^- \to e^- e^-$ decay in a muonic atom offers
significant advantages. From an experimental point of view, 
and in addition to having a larger phase space, it leads to a cleaner
experimental signature than the 3-body $\mu^+ \to e^+ e^- e^-$ decay;
the LFV muonic atom decay has a two-body final state in which the
electrons are emitted nearly back-to-back, each with a well-defined
energy\footnote{Neglecting variations due to bound effects of the 1s
  states as well as the Coulomb interaction from the
  nucleus~\cite{Koike:2010xr}.},  
$E_{e^-}\sim m_\mu/2$. 
Furthermore, the rate of this process can be enhanced due to the
Coulomb attraction from the nucleus, which increases the overlap of
the 1s electron and muon wavefunctions,
$\psi^{(e),(\mu)}_\text{1s}$. Following~\cite{Koike:2010xr}, one can write
the transition rate of the $\mu^- e^- \to e^- e^-$ process as
\begin{eqnarray}\label{eq:cross.section:vrel.psi}
\Gamma(\mu^- e^- \to e^- e^- , \text{ N})\, &=& \, \sigma_{\mu e \to
  ee} v_\text{rel} \, 
|\psi_\text{1s}^{(e)} (0; Z-1)|^2\, , \quad \nonumber\\
 \text{with } 
\quad\psi_\text{1s}^{(e)} (0; Z-1)\, &=&\, 
\frac{\left[ (Z-1)\, \alpha\, m_e \right]^{3/2} }{\sqrt \pi} ,
\end{eqnarray}
where $\sigma_{\mu e \to ee} v_\text{rel}$ denotes the electroweak
cross-section.
Thus, and when compared to the (apparently) similar Muonium decay  
$\mu^+ e^- \to e^+ e^-$, the effects of the Coulomb interaction 
lead to an enhancement of the muonic atom decay rate by 
a factor $\sim (Z-1)^3$, which can become important for nuclei with
large atomic numbers. 

The most recent results of~\cite{Uesaka:2015qaa}
emphasised the importance of taking into account the distortion effect of 
the emitted electrons due to the nuclear Coulomb potential, and of
performing a relativistic treatment of the wave function of the bound
leptons.
Whilst for small atoms the enhancement is indeed well described
by $f_\text{Coul.}(Z) \approx (Z-1)^3$, for large $Z$ atoms the rate can be
enhanced by as much as an additional order of magnitude (see  
Fig.~1 of~\cite{Uesaka:2015qaa}).  
In our study we take into account these results (in particular we use
the data obtained for a uniform distribution of the nuclear charge).

As was the case for the coherent conversion in nuclei, 
in extensions of the SM via sterile fermions, the effective Lagrangian
describing the muonic atom LFV decay contains $\gamma$-dipole
(long-range) interactions and ``local'' (contact)
terms~\cite{Kuno:1999jp,Koike:2010xr}.
The contribution to the cross-section of 
the (long-range) photonic interactions is subdominant for the present
class of NP models\footnote{We have numerically verified that
  this assumption is indeed correct for approximately all regions in
  the parameter spaces of the different models under consideration.}; 
moreover, in the absence of a complete
estimation of the associated nuclear effects~\cite{Uesaka:toappear,
  Kuno:private}, we will not take into account the ``long-range''
$\gamma$-dipole contributions. 
The contact interactions include contributions from box diagrams, as
well as photon 
and $Z$ penguin diagrams; in a muonic atom, with an atomic number $Z$,
they contribute to the branching ratio of the process in
Eq.~(\ref{eq:me2ee}) as
\begin{align}
&\text{BR}(\mu^- e^- \to e^- e^- , \text{ N})\, \equiv 
\, \tilde{\tau}_{\mu}\,
  \Gamma(\mu^- e^- \to e^- e^- , \text {N}) \nonumber \\ 
& = 24\pi \, f_\text{Coul.}(Z)\, 
\alpha_w  \left( \frac{m_e}{m_{\mu}} \right)^{3}\,
  \frac{\tilde{\tau}_{\mu}}{\tau_{\mu}}\,
  \left(16 \, \left | \frac{1}{2} \left (\frac{g_w}{4 \pi} \right)^2
  \left (\frac{1}{2}F^{\mu eee}_\text{Box} + F_Z^{\mu e} - 
  2 \sin^2\theta_w \left (F_Z^{\mu e} - F_\gamma^{\mu e} \right)
  \right) \right|^2 + \right . \nonumber\\ 
& \left . +  4  \, \left| \frac{1}{2} \left(\frac{g_w}{4 \pi} \right)^2 
  2 \sin^2\theta_w \left (F_Z^{\mu e} - F_\gamma^{\mu e} \right)
 \right|^2 \right) \, , 
\label{eq:br-4fermi}
\end{align}
where we have neglected the interference between contact terms
(which can be sensitive to CP violating phases that might emerge in
relation to flavour violating
effects).  In the above equation 
$\tilde{\tau}_{\mu}$ corresponds to the lifetime of a muonic atom,
which depends on the specific element, and is always smaller 
than the lifetime of free muons, $\tau_{\mu}$ 
($\tau_{\mu}= 2.197 \times 10^{-6}$~s~\cite{Agashe:2014kda}). 
In Table~\ref{table:muonic.atom:lifetime} we summarise the values of
$\tilde{\tau}_{\mu}$ 
for different elements~\cite{Suzuki:1987jf}, 
which will be relevant in our discussion. 
\renewcommand{\arraystretch}{1.3}
\begin{table}[h!]
\centering
\begin{tabular}{|c|c|}
\hline
Element & $\tilde{\tau}_{\mu}$ (s)\\
\hline
$^{1}$H & $2.19 \times 10^{-6}$ \\
$^{13}$Al & $8.64 \times 10^{-7}$\\
$^{79}$Au & $7.26 \times 10^{-8}$\\
$^{92}$U & $7.5 \times 10^{-8}$\\
\hline
\end{tabular}
\caption{Lifetime of a muonic atoms: Hydrogen, Aluminium, Gold and
  Uranium~\cite{Suzuki:1987jf}.} 
\label{table:muonic.atom:lifetime}
\end{table}
\renewcommand{\arraystretch}{1.}
In Eq.~(\ref{eq:br-4fermi}), $F_{\gamma,Z}^{\mu e}$ 
denote the contributions from photon- and $Z$-penguins
(corresponding to those already introduced in Eq.~(\ref{eq:CRmue}) 
of the previous subsection); $F^{\mu eee}_\text{Box}$ is defined in
Appendix~\ref{app:A}. 
As can be inferred from the nature of the contributing diagrams, 
the muonic atom $\mu^- e^- \to e^- e^-$ decay 
arises from the same elementary processes as the rare 3-body decay
$\mu^+ \to e^+ e^- e^-$. Together with the bounds arising from the
coherent conversion in Nuclei (CR($\mu-e$, N)), the present
experimental bounds on BR($\mu \to e e e$) may indirectly constrain the
possible values of BR($\mu^- e^- \to e^- e^-$, N). Thus, and for
completeness, we collect the relevant expressions for 
the 3-body muon decay in Appendix~\ref{app:B}.

As mentioned in the Introduction, the $\mu^- e^- \to e^- e^-$ process
could be investigated 
by the COMET collaboration (possibly being part of its Phase II 
programme). In our
numerical analysis, we will thus work under the hypothesis of a similar
future sensitivity for BR($\mu^- e^- \to e^- e^-$, Al) and CR($\mu-e$,
Al).

\subsection{Muonium:  Mu-$\rm \bf \overline{Mu}$
  conversion and decay}\label{sec:cLFV:muonic:mubarmu} 

The Muonium (Mu) atom is a Coulomb  
bound state of an electron and an anti-muon
($e^-\mu^+$)~\cite{Pontecorvo:1957cp}, formed when a $\mu^+$ slows
down inside matter and captures an $e^-$. 
This hydrogen-like state is
actually free of hadronic interactions, and its electromagnetic 
binding is well described by the SM electroweak interactions. In turn,
this renders the Muonium
system an interesting laboratory to accurately determine
fundamental constants, or test for deviations from the SM
induced by the presence
of possible NP states and interactions. 

The spontaneous conversion of a Muonium atom to its anti-atom 
($\overline{\text{Mu}} = e^+\mu^-$)
has been identified as 
an interesting class of muon cLFV process~\cite{Feinberg:1961zza}. 
Similar to the other cLFV
observables, the observation of Mu-$\overline{\text{Mu}}$ (also
known as Muonium-antimuonium oscillation)
would constitute a clear signal of
physics beyond the SM. 

The nontrivial mixing between the two states 
comes from the non-vanishing LFV transition amplitude for $e^-
\mu^+ \to e^+ \mu^-$, which corresponds to the simultaneous violation
of individual electron and muon numbers, $|\Delta L_{e,\mu}|=2$.  
Under the assumption\footnote{There
  could be in general various possible combinations of four-fermion
  interactions at the origin of the Mu-$\rm \overline{Mu}$ transition,
  involving vector, axial, scalar and pseudoscalar effective
  interactions. Different NP models (adding new particles
  and/or interactions) could lead to such contributions. For a
  detailed discussion  see~\cite{Kuno:1999jp} and references therein. 
} of $(V -A) \times (V - A)$ interactions, the 
Mu-$\rm \overline{Mu}$ transition can be described by an effective
four-fermion interaction with a coupling constant $G_{\rm M\overline{M}}$, 
\begin{equation}
 \mathcal{L}_\text{eff}^{\rm M\overline{M}} \,= \, \frac{G_{\rm
     M\overline{M}}}{ \sqrt{2} } 
\left[\, {\overline \mu}\,  \gamma^{\alpha} (1 - \gamma_5) \,e
\,\right] \left[ \,{\overline \mu}\, \gamma_{\alpha} (1 - \gamma_5)\,
e \,\right] \, .
\label{eq:Leff_muonium}
\end{equation}
It is worth noticing that the new interactions will cause a splitting
of the energy levels of Muonium and antimuonium (which, in the absence
of an external magnetic field, would otherwise have degenerate ground
state energy levels). This splitting can be also cast in terms of the
effective coupling as~\cite{Kuno:1999jp} 
\begin{equation}\label{eq:mu-barmu:Esplit}
\delta_E^{\text{M}-\overline{\text{M}}} \, = \, 
\frac{8 \, G_F}{\sqrt 2 \, n^2 \pi a_0^3}\, \left(
\frac{G_{\rm M\overline{M}}}{G_F} \right)\,,
\end{equation}
where $n$ and $a_0$ denote the principal quantum number and the Bohr
radius of the muonic atom. 

Searches for  Mu-$\rm \overline{Mu}$ conversion were started more than
fourty years ago, and have employed different techniques. At present,
no positive signal has been found;
the best limit has been set at PSI~\cite{Willmann:1998gd}, where 
Muonium atoms are formed by electron capture when positively charged 
muons (from a very intense muon beam) are stopped in a SiO$_2$ powder
target. 
Experimental searches put a bound on $\left |\text{Re}\left( G_{\rm M\overline{M}} \right) \right|$, or on its
ratio to $G_F$, and the limit(s) are usually determined  
assuming that the undelying interaction is of the type $(V \pm A)
\times (V \pm A)$; for these cases, the 90\%C.L. limit on the
effective coupling constant has been reported to be 
$\left |\text{Re}\left( G_{\rm M\overline{M}} \right) \right| \leq 3.0 \times 10^{-3}G_F$~\cite{Willmann:1998gd}. 

On Table~\ref{table:muonium:bounds} we summarise the different bounds 
so far obtained for the conversion probability P(Mu-${\rm \overline{Mu}}$),  
translated into upper bounds on the effective coupling $G_{\rm
  M\overline{M}}$. With the advent of new, very intense muon sources, it
can be expected that the current bounds will be improved in the near
future. 
\begin{table}[tb!]
\centering
\begin{tabular}{|c|c|c|}
\hline
Experiment & P(Mu-${\rm \overline{Mu}}$) & $\left |\text{Re}\left( G_{\rm M\overline{M}} \right) \right| /G_F$
\\ 
\hline
Huber et al. (1990) \cite{Huber:1989by} & 
$< 2.1 \times 10^{-6}$ &  $< 0.29$ \\
Matthias et al. (1991) \cite{Matthias:1991fw} & 
$< 6.5 \times 10^{-7}$ &  $< 0.16$ \\
Abela et al. (1996) \cite{Abela:1996dm} & 
$< 8.0 \times 10^{-9}$ &  $< 0.018$ \\
Willmann et al. (1999) \cite{Willmann:1998gd} & 
$< 8.3 \times 10^{-11}$ &  $< 0.003$ \\
\hline
\end{tabular}
\caption{Experimental results (90\% C.L.) for the
    conversion probability and the corresponding upper bound for the
    coupling  $G_{\rm M\overline{M}}$. }
\label{table:muonium:bounds}
\end{table}

\begin{figure}
\begin{center}
\epsfig{file=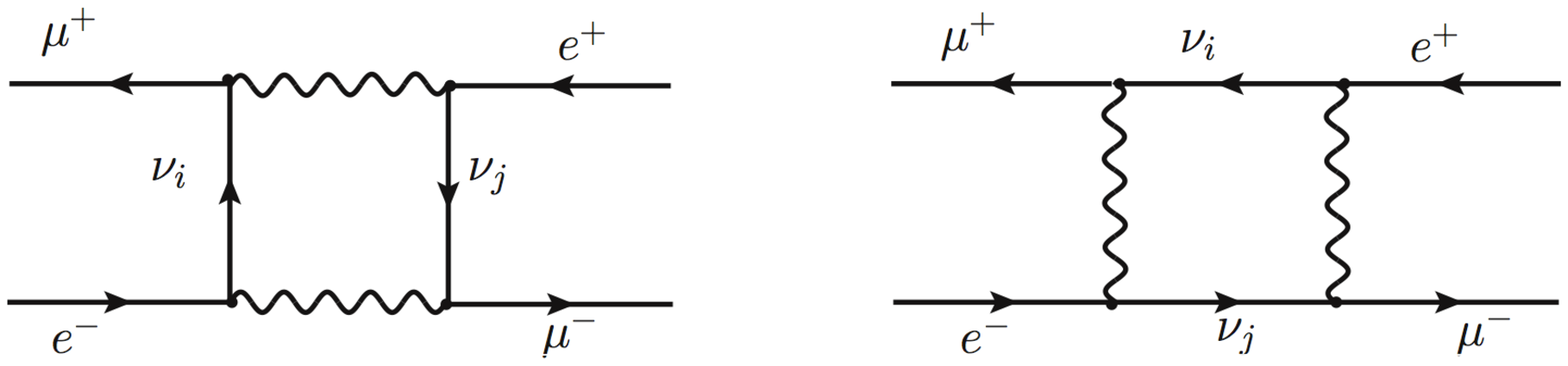,width=115mm} 
\epsfig{file=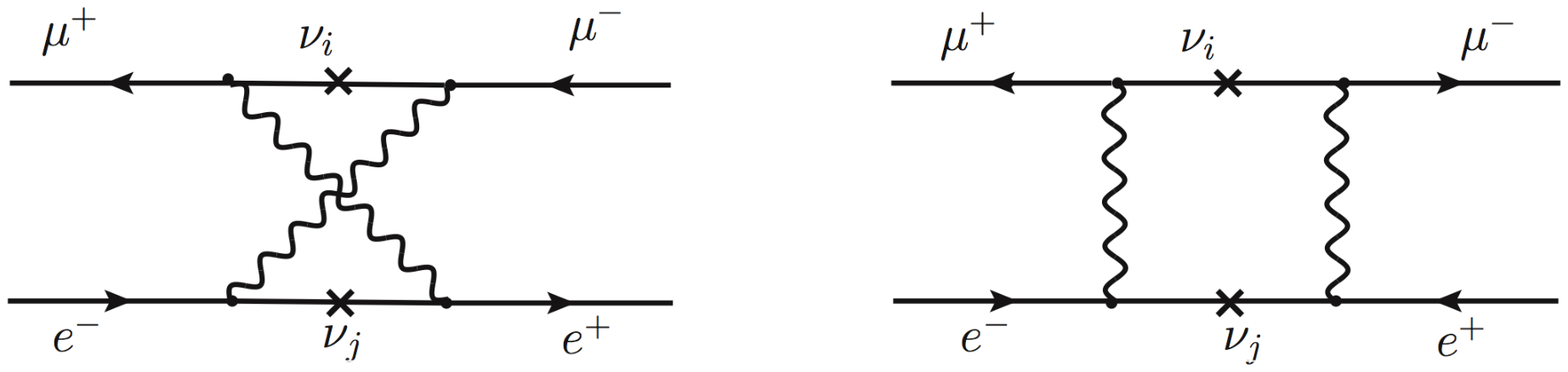,width=115mm} 
\end{center}
\caption{Muonium-antimuonium conversion: 
  box diagrams for generic (Dirac or Majorana) neutrinos (upper
  boxes) and box diagrams exclusively for Majorana neutrinos (below). The
  wavy lines stand either for a $W^\pm$ or a charged Goldstone boson. } 
\label{fig:muonium_boxes}
\end{figure}

In extensions of the SM with sterile neutrinos, the $e^-
\mu^+ \to e^+ \mu^-$ transition responsible for  Mu-$\rm
\overline{Mu}$ conversion occurs at the loop-level via box diagrams. 
There are four different diagrams which can be separated in two
contributions, those which are common to both Dirac and Majorana
neutrinos (Fig.~\ref{fig:muonium_boxes}, upper boxes) and two other which
appear only if neutrinos are Majorana particles
(Fig.~\ref{fig:muonium_boxes}, lower boxes). 
The computation of the box diagrams of Fig.~\ref{fig:muonium_boxes}
(in a unitary gauge) allows to write the 
effective coupling $G_{\rm M\overline{M}}$ 
as~\cite{Clark:2003tv,Cvetic:2005gx}:
\begin{equation}
\frac{G_{\rm M\overline{M}}}{\sqrt{2}}\,=\,
-\frac{G_F^2 M_W^2}{16\pi^2}\left[\sum_{i,j =1}^{3+n_S}
({\bf U}_{\mu i}\,{\bf U}^\dagger_{e i})\,({\bf U}_{\mu j}\,
{\bf U}^\dagger_{e j})\,{\rm G}_{\rm Muonium}(x_i, x_j)\right]
\label{eq:Gmumu} ~,
\end{equation}
where $x_i =\frac{m_{\nu_i}^2}{M_W^2}, i=1,...,3+n_S$ and ${\rm
  G}_{\rm Muonium}(x_i, x_j)$ is the loop 
function arising from the two groups of boxes (generic
and Majorana), and is given in Appendix~\ref{app:A}. 

This observable has been addressed in the context of several
extensions of the SM via additional sterile states, mostly in relation with
type I seesaw realisations with heavy RH 
neutrinos~\cite{Clark:2003tv,Cvetic:2005gx,Liu:2008it}.  

\bigskip
Although a Muonium state mostly decays via standard channels (the most
frequent one  
being $\text{Mu} \to e^+ e^- \bar \nu_\mu \nu_e$, or equivalently, a
SM muon decay with the electron as a spectator), 
the presence of NP can also induce the cLFV decay 
\begin{equation}\label{eq:Mudecay:def} 
\text{Mu} \, \to \, e^+ \, e^-\, . 
\end{equation}
In SM extensions with RH neutrinos, 
these decays have been addressed at low-energies, and also in the
framework of a future muon collider~\cite{Cvetic:2006yg}. 
The cLFV Muonium decay rate can be written as 
\begin{equation}\label{eq:Mudecay:BR}
\text{BR}(\text{Mu} \to e^+ e^-) \, =\, 
\frac{\alpha^3}{\Gamma_\mu \, 32 \pi^2}\, 
\frac{m_e^2 m_\mu^2}{(m_e + m_\mu)^3}\,
\sqrt{1 -4\, \frac{m_e^2}{(m_e + m_\mu)^2}}\,
|\mathcal{M}_\text{tot}|^2\, 
\end{equation}
with $\Gamma_\mu$ the muon decay width,
and where $|\mathcal{M}_\text{tot}|$ denotes the full amplitude, summed
(averaged) over final (initial) spins~\cite{Cvetic:2006yg}.
The full expression for $|\mathcal{M}_\text{tot}|$ is given in
Appendix~\ref{app:A}; we notice that the intervening quantities 
are those also present for 3-body
lepton decays (although the distinct form factors - penguins,
long-range dipoles, boxes - contribute
differently to the matrix element). 

The experimental roadmap concerning this last cLFV observable is not
clear; we will thus make no reference to experimental limits or
sensitivities in the numerical analysis. 

\section{cLFV in muonic atoms: numerical results} 
\label{sec:clfv.results} 
We now investigate the cLFV observables presented in the previous 
section, for  different
classes of SM extensions: we first 
carry a detailed analysis for a simple toy model (``3+1''), and then
highlight the most important points emerging from the study of 
two well motivated NP models, the ISS and the $\nu$MSM. 

\subsection{``3+1 model'' toy model: nuclear-assisted cLFV
  processes}\label{sec:res:3+1} 
As described in Section~\ref{sec:steriles:th}, this simple model
allows to illustrate the potential effects of the addition of $n_S$
sterile fermion states, ``encoded'' in the contribution of $\nu_4$.

\subsubsection{$\pmb{\mu \to e}$ conversion in nuclei}\label{sec:res:3+1:CR}
\begin{figure}
\begin{center}
\epsfig{file=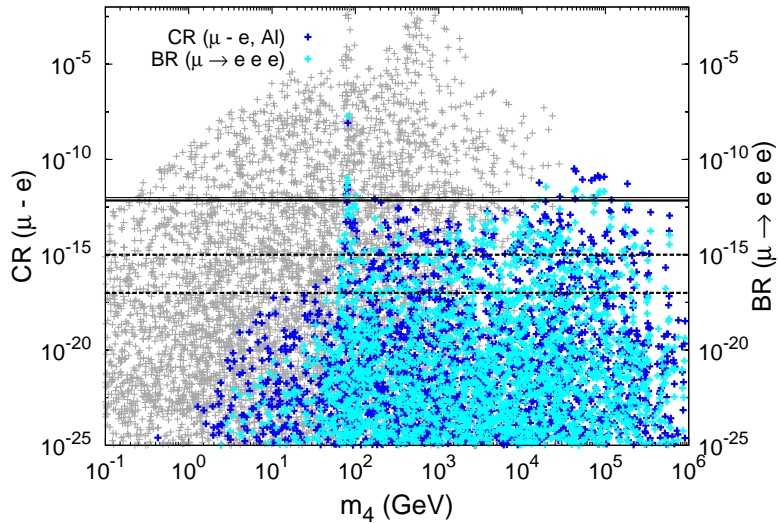, width=75mm,
  angle=270} 
\end{center}
\caption{Effective ``3+1 model'': CR($\mu -e$, Al) and BR($\mu \to e e
  e$) as a function of the mass of the mostly sterile state $m_4$. 
The former is displayed in dark blue (left axis), while the latter is
depicted in cyan (right axis). 
Grey points correspond to the violation of at least
one experimental bound (other than those arising from CR($\mu -e$, Au)
and BR($\mu \to e e e$)). A thick (thin) solid horizontal line denotes
the current experimental bound on the CR($\mu -e$,
Au)~\cite{Bertl:2006up}
 ($\mu \to e e e$ decays~\cite{Bellgardt:1987du}), 
while dashed lines correspond to future sensitivities to 
CR($\mu -e$, N)~\cite{Alekou:2013eta,Kuno:2013mha}, see
Tables~\ref{table:mue:bounds.past} and \ref{table:mue:future}. } 
\label{fig:EFF_CRmue.mu3e_m4}
\end{figure}

We begin our study by considering the impact of the additional sterile
state concerning the coherent $\mu \to e$ conversion in a muonic
atom. We choose to illustrate this observable for the specific case of
an Aluminium nucleus; similar results would be obtained for other
nuclei (as Gold or Titanium)\footnote{As discussed
  in~\cite{Kitano:2002mt}, the 
  theoretical predictions for typical NP models (arising either from
  long-range photonic dipoles, scalar or vector contact operators) 
  vary by as little as a 1.5 - 2 factor between Aluminium, Titanium
  and Gold nuclei.}.
The results of a comprehensive scan over the additional degrees
of freedom of the sterile state
(as described in Section~\ref{sec:steriles:th}) are displayed in
Fig.~\ref{fig:EFF_CRmue.mu3e_m4} in which, for completeness, we
also include the predictions of the simple ``3+1 toy model'' for the
three body $\mu \to e e e$ decay. In grey we denote points violating at
least one of the experimental constraints listed in
Section~\ref{sec:steriles:constraints}, with the exception of the
observables under study.    
Coloured points denote the two observables: in dark blue the
predictions for CR($\mu -e$, Al), corresponding to the left $y$-axis; 
in cyan, the values of BR($\mu \to e e e$), displayed on the right 
$y$-axis. (Cosmological bounds would have typically 
disfavoured regions in parameter space for which $m_4 \lesssim
0.1$~GeV, and hence are not visible in Fig.~\ref{fig:EFF_CRmue.mu3e_m4}.)  

Minimal extensions of the SM via sterile fermions 
- such as the simple ``3+1 toy model'' -
can easily account for sizable contributions to both CR($\mu -e$, Al)
and BR($\mu \to e e e$), even 
above the current experimental bounds
(respectively depicted by thick and thin solid horizontal lines in 
Fig.~\ref{fig:EFF_CRmue.mu3e_m4}). 
This is in good agreement with
previous analyses carried for specific models (e.g. low-scale type I
seesaw~\cite{Alonso:2012ji}, ISS~\cite{Abada:2014kba}). 
 For the mass regime of the
mostly sterile state around the EW scale ($m_4 \sim 10^2$ GeV), 
the leading contributions arise from $Z$-penguin
diagrams; below the EW scale, box diagrams become
increasingly important, and dominate the total width below a few~GeV. 
(For the low mass regime, points are excluded as neutrino data
cannot be accommodated, mostly due to an excessive departure from
unitarity of the $\tilde U_\text{PMNS}$; bounds from neutrinoless
double beta decays are also important in this mass regime.)
For very large values of the mostly sterile heavy neutrino mass, 
$m_4 \gtrsim 10$~TeV, the bound arising from CR($\mu -e$, Au) becomes
the most constraining one. 

As mentioned before, the scan leading to the results displayed in 
Fig.~\ref{fig:EFF_CRmue.mu3e_m4} explores all the new degrees of
freedom of the additional sterile state, in particular the possible
extra CP violating phases. We have verified that, as expected, these
do not  play a significant r\^ole for this type of observables. We
have also considered an IH spectrum for the light neutrino spectrum,
finding that this leads to similar results concerning the cLFV observables
(the only significant difference being that bounds from neutrinoless double beta
decay now exclude more important regions of the parameter space).

The coherent $\mu-e$ conversion in muonic atoms, induced by an
additional sterile neutrino, could certainly be probed in near future
experiments, as Mu2e or COMET (both with Aluminium targets). 
In fact, for masses of the heavy (mostly) 
sterile state above the EW scale, the
predictions of this simple model are well within reach 
 of the first COMET phase - whose
sensitivity corresponds to the upper horizontal dashed line 
in Fig.~\ref{fig:EFF_CRmue.mu3e_m4} (see
Table~\ref{table:mue:future}) - or even DeeMe, which is 
expected to reach an
$\mathcal{O}(10^{-14})$ sensitivity, albeit for a silicon-carbide target. 

\subsubsection{Decay of muonic atoms to $e^- e^-$
  pairs}\label{sec:res:3+1:mue2e} 
Despite the apparent similarity to the $\mu \to e e e$ decay, the
(Coulomb enhanced) decay of a muonic atom to a pair of electrons
substantially differs from the 3-body decay, 
both at the theoretical and at the experimental
level. 

\begin{figure}
\begin{center}
\epsfig{file=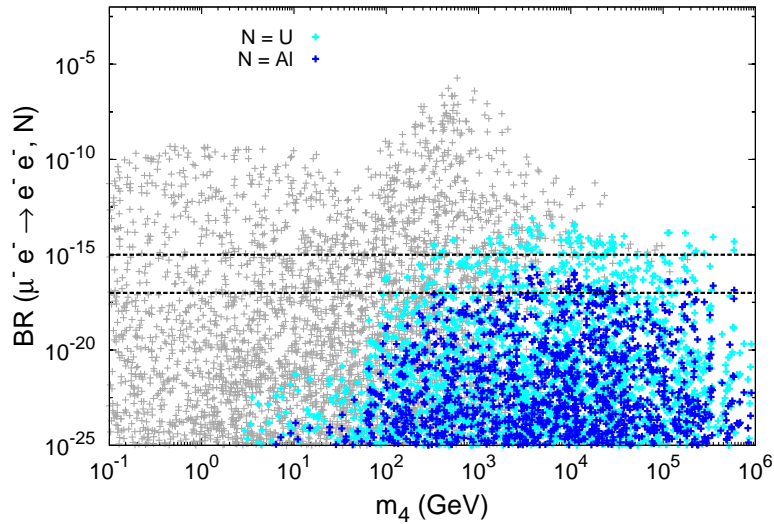, width=75mm,
  angle=270}    
\end{center}
\caption{Effective ``3+1 model'': BR($\mu^- e^- \to e^- e^-$, N) 
as a function of the mass of the mostly sterile state $m_4$, for two
distinct muonic atoms, Aluminium (dark blue) and Uranium (cyan). 
Grey points correspond to the violation of at least
one experimental bound;  
dashed horizontal lines denote the future sensitivity of COMET (Phase I and
II)~\cite{Kuno:2013mha}. }
\label{fig:EFF_mue2ee_AlU_m4}
\end{figure}

Firstly, and as emphasised in the discussion of
Section~\ref{sec:cLFV:muonic:mue2e},  
the process' rate can be significantly enhanced in large $Z$ atoms (in
particular the contributions from contact
interactions~\cite{Uesaka:2015qaa}). We thus begin
the numerical analysis by comparing the prospects for two different
nuclei; this is illustrated in Fig.~\ref{fig:EFF_mue2ee_AlU_m4} for  
Aluminium ($Z=13$, dark blue) and Uranium ($Z=92$, cyan). Grey points
correspond to the violation of at least one experimental bound: the
most stringent constraints arise, as expected, from $\mu \to e e e$ (and
also from CR($\mu - e$, Au)). 

The Coulomb
enhancement is clearly visible: should this process be included in
COMET's physics programme, the cLFV muonic atom decay should be within reach
of COMET's Phase II, even for light nuclei, such as Aluminium 
(in the regime 
$m_4 \gtrsim 200$~GeV); for heavier
atoms, such as Uranium, branching ratios above $10^{-15}$ render this
process experimentally accessible (a similar situation occurs for
Lead nuclei - albeit suppressed by a factor $\sim
7/9$ when compared to Uranium~\cite{Uesaka:2015qaa}). 

As mentioned in Section~\ref{sec:cLFV:muonic:mue2e}, in the absence of
a complete 
estimation of the nuclear effects regarding the long-range photon cLFV
interaction, we only considered contributions from
contact interactions. It is possible that the additional dipole
interactions further increase the total BR($\mu^- e^- \to e^-
e^-$, N)~\cite{Uesaka:2015qaa,Uesaka:toappear}.  

\begin{figure}
\begin{center}
\epsfig{file=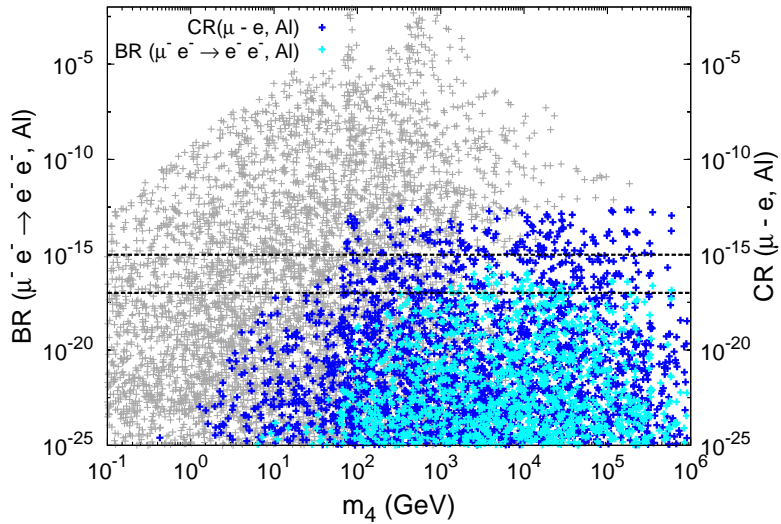,
  width=75mm, angle=270} 
\end{center}
\caption{Effective ``3+1 model'': BR($\mu^- e^- \to e^- e^-$, Al)
  (cyan, left axis) and 
CR($\mu -e$, Al) (dark blue, right axis) as a function of the mass of the 
mostly sterile state $m_4$.  
Grey points correspond to the violation of at least
one experimental bound;  
dashed horizontal lines denote the future sensitivity of COMET (Phase I and
II)~\cite{Kuno:2013mha}. }
\label{fig:EFF_CR.mue2ee_m4}
\end{figure}

\bigskip
Should the decay of the muonic atom ($\mu^- e^- \to e^- e^-$) be
included in the physics programme of dedicated high-intensity
facilities (as COMET), then it is only natural to investigate to which
extent it can become a powerful probe of cLFV, and how it can
complement information obtained from other nuclear assisted processes,
such as $\mu - e $ conversion.  We thus display in
Fig.~\ref{fig:EFF_CR.mue2ee_m4} the expectations for both observables, 
BR($\mu^- e^- \to e^- e^-$, Al) and CR($\mu - e$, Al) 
in the simple ``3+1'' toy model. 
In general, the coherent conversion appears to have a
stronger experimental potential, with contributions well within COMET 
reach for sterile masses above a few tenths of GeV. Interestingly, 
for the low-mass regime, the sterile fermion contributions to the
muonic atom decay (via the
box diagrams, as already discussed before) could be 
much
larger, but this regime is
heavily constrained on theoretical and experimental 
arguments. Heavier atoms, such
as Lead, would further enhance these contributions.   

\begin{figure}
\begin{center}
\epsfig{file=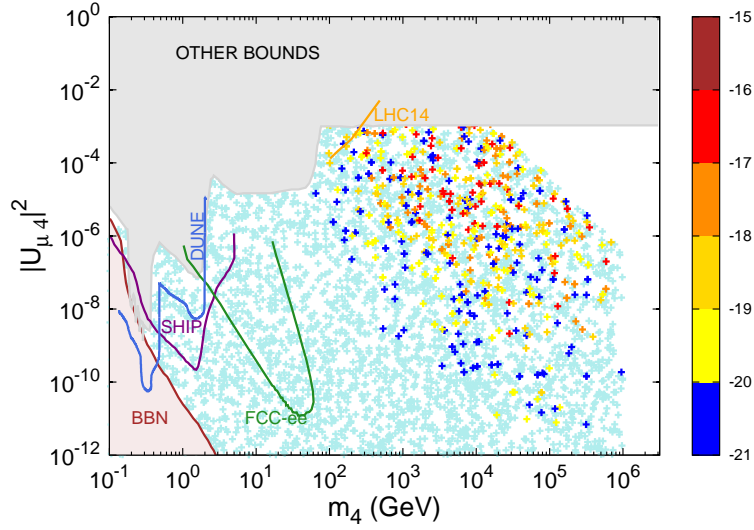, width=75mm,
  angle=270} 
\end{center}
\caption{Effective ``3+1 model'': $(|U_{\mu 4}|^2, m_4)$ parameter
  space. The shaded surfaces correspond to the exclusion from BBN
  (rose) or from the violation of at least one experimental
  bound (grey). Solid lines delimit the expected sensitivity of
  several facilities: DUNE (blue), SHiP (violet), FCC-ee (green) and
  LHC (orange). The coloured points denote the predictions for the  
  logarithm of BR($\mu^- e^- \to e^- e^-$, Al):    
  from larger to smaller values (dark red, yellow, blue). 
  Cyan denotes values of the branching fraction below $10^{-21}$.
} 
\label{fig:EFF_U2m4.palette}
\end{figure}

To conclude the discussion of this observable, we present in
Fig.~\ref{fig:EFF_U2m4.palette} the predictions for the logarithm of
the BR($\mu^- e^- \to e^- e^-$, Al) in the 
$(|U_{\mu 4}|^2, m_4)$ 
parameter space of the simple ``toy model". Shaded surfaces reflect the
violation of at least one phenomenological or cosmological bound 
(see Section~\ref{sec:steriles:constraints}, as well 
as~\cite{Alekhin:2015byh},~\cite{Deppisch:2015qwa}). 
The different solid lines correspond to the reach of future
facilities: the projected exclusion limit from the LHC (14~TeV centre of
mass energy, with 300~fb$^{-1}$ data~\cite{Deppisch:2015qwa}); the
expected sensitivity of FCC-ee regarding the production of heavy (RH)
sterile neutrinos from $Z \to \nu_\ell \nu_s$ 
(estimated $10^{12}$ $Z$ decays, for a 10-100 cm decay
length~\cite{Blondel:2014bra}); DUNE (former LBNE, a beam dump
experiment, searching for the decay products of sterile neutrinos
produced in charmed meson decays)~\cite{Adams:2013qkq}; SHiP (a 
fixed-target experiment using high-intensity proton beams at the CERN
SPS~\cite{Bonivento:2013jag,Anelli:2015pba}). 
As can be seen, there is little overlap between points 
associated with BR($\mu^- e^- \to e^- e^-$, N) within COMET sensitivity and the
reach of the above mentioned facilities. This is mostly a consequence
of the large mass regime responsible for the latter contributions,
which precludes the (direct) production of the sterile states.  

The potential sensitivity of COMET concerning this observable (which explores
cLFV in the $\mu-e$ sector) would be complementary to that of 
cLFV in the $\mu-\tau$ sector, in particular concerning high-energy
observables such as the decay $Z\to \mu \tau$, which could be studied at 
FCC-ee (TLEP), running in $e^+ e^-$ mode close to the $Z$ mass
threshold. In fact, most of the points within
  COMET sensitivity are predicted to account for a 
  BR($Z\to \mu \tau$) lying within FCC-ee
  reach ($\gtrsim 3 \times 10^{-13}$~)\cite{Abada:2014cca}. 
  This implies that this simple  
  ``3+1'' toy model can be probed via two fully independent - and yet
  strongly complementary - cLFV experimental approaches.

\subsubsection{Muonium oscillation and decays}\label{sec:res:3+1:mubarmu}
As discussed in Section~\ref{sec:cLFV:muonic:mubarmu}, Muonium
consists of a hydrogen-like atom (although free of hadronic
interactions). 
We proceed to summarise the prospects regarding
the sterile neutrino contributions to Muonium-antimuonium transitions
and Muonium decay.

\begin{figure}
\hspace*{-11mm}
\begin{tabular}{cc}
\epsfig{file=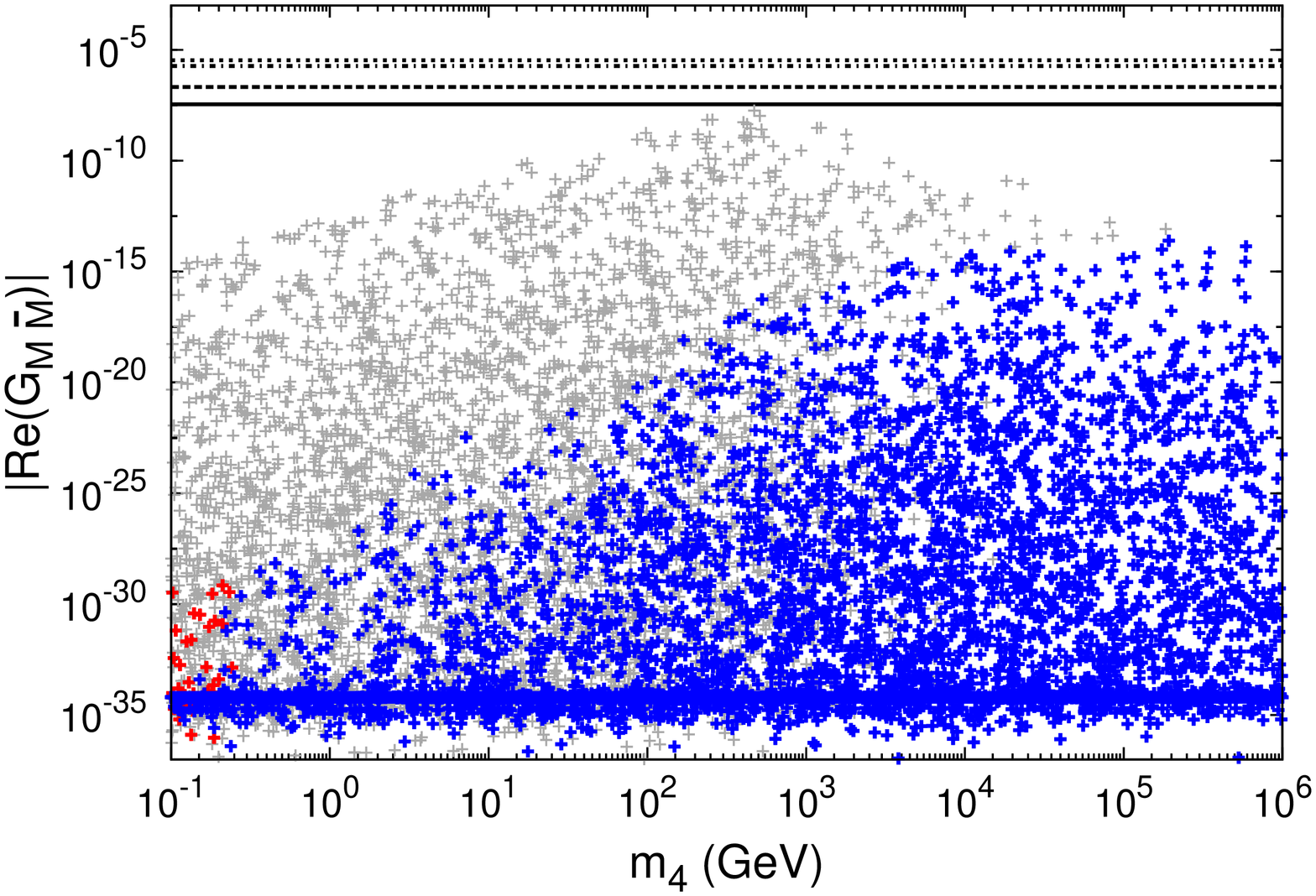,width=63mm,
  angle=270} \hspace*{-3mm}
&\hspace*{-3mm}
\epsfig{file=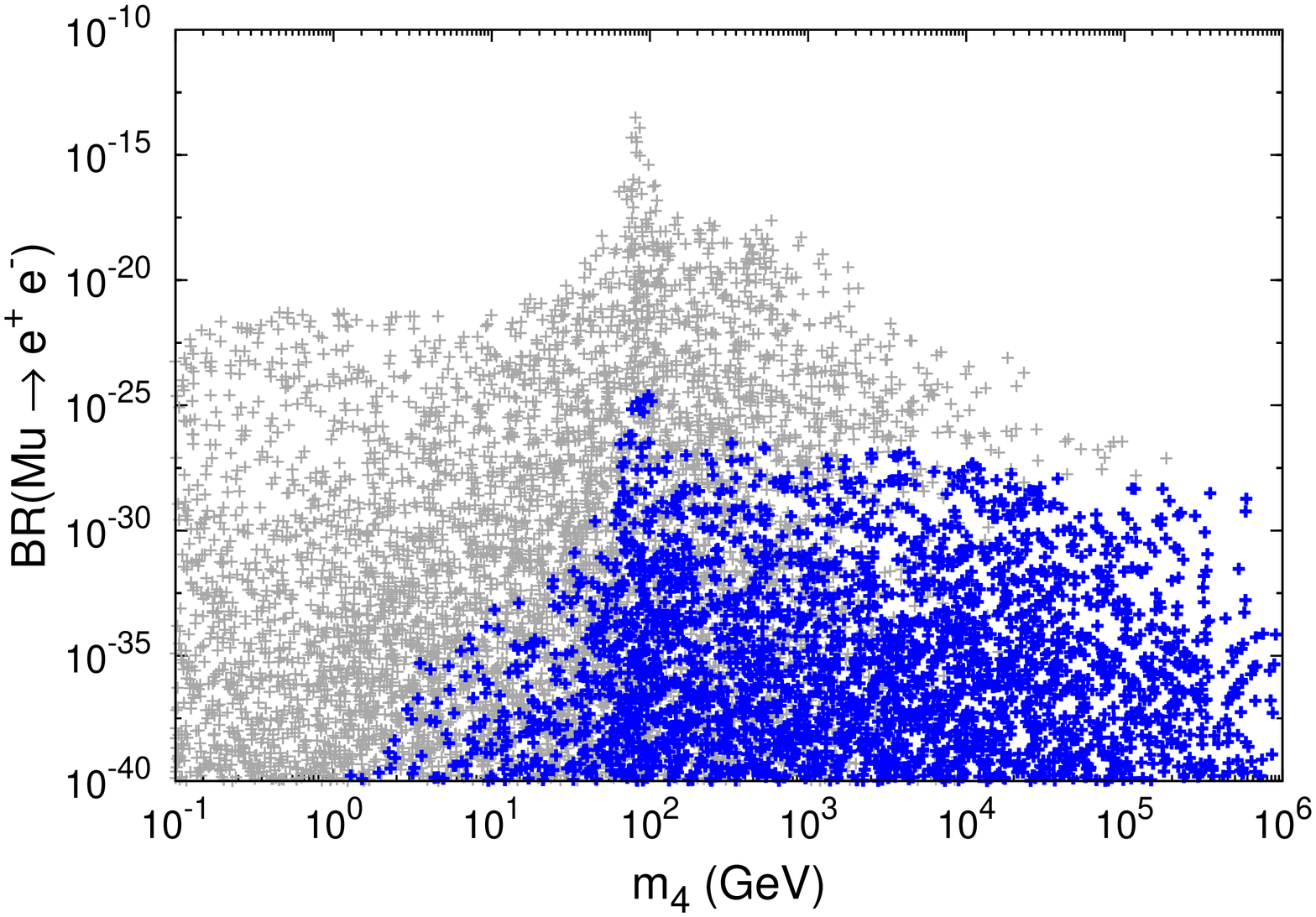,width=63mm,
  angle=270}  
\end{tabular}
\caption{Effective ``3+1'' model:  effective coupling $G_\text{M$\rm
    \overline{M}$}$ ($\left
|\text{Re}\left(G_{\rm M\overline{M}} \right) \right|$) for Mu - $\rm \overline{Mu}$ conversion
 (left panel) and  
cLFV decay rate BR$(\text{Mu} \rightarrow e^+ e^-)$ (right panel), 
as a function of the mass of the mostly sterile state $m_4$. 
Grey points correspond to the violation of at least
one experimental bound; red points illustrate regimes which are 
disfavoured from standard cosmology bounds, while dark blue points are
in agreement will all available bounds. In the left panel, the
horizontal lines denote 
the evolution of the experimental bounds and constraints, from older
(dotted), to the most recent one (full), 
see Table~\ref{table:muonium:bounds}.} 
\label{fig:EFF_muonium_m4}
\end{figure}

Figure~\ref{fig:EFF_muonium_m4} displays the expected contributions
for the simple ``3+1'' toy model, both to the 
Mu - $\rm \overline{Mu}$ conversion (left) and to the 
cLFV Muonium decay  
$\text{Mu} \rightarrow e^+ e^-$ (right). 
For completeness, the colour scheme of this figure illustrates the
regimes which would be excluded from violation of cosmological bounds
(red). 

Although the sterile contribution could potentially account for values
of the effective coupling constant not too far from the most recent
measurement, these points are excluded as they would be associated to
excessive CR($\mu -e$, Au) and BR($\mu \to e e e$). The
phenomenologically viable 
parameter space can lead to maximal values 
$\left |\text{Re}\left(G_{\rm M\overline{M}} \right) \right| \sim 10^{-13}$, with a saturation (pure
$U_\text{PMNS}$-like) at around $10^{-25}$; likewise\footnote{Our
  results are in agreement with the findings of~\cite{Cvetic:2006yg}
  (in the appropriate limits).}, one expects
maximal values of BR($\text{Mu}
\rightarrow e^+ e^-$)~$\sim 10^{-25}$.

Contrary to the previous observables, for which the r\^ole of these SM
extensions would be to accommodate (or explain) possible experimental 
signals, the potential observation of Mu - $\rm \overline{Mu}$
conversion in the (near) future would strongly disfavour such a minimal
extension.

\subsection{Low-energy seesaw models}\label{sec:res:others}
To conclude our discussion, we illustrate the contributions to the
``nuclear assisted'' cLFV observables arising in the framework of two
well-motivated low-energy seesaw models. 

\subsubsection{The (3,3) Inverse Seesaw realisation}\label{sec:res:others:ISS}
\begin{figure}
\begin{center}
\epsfig{file=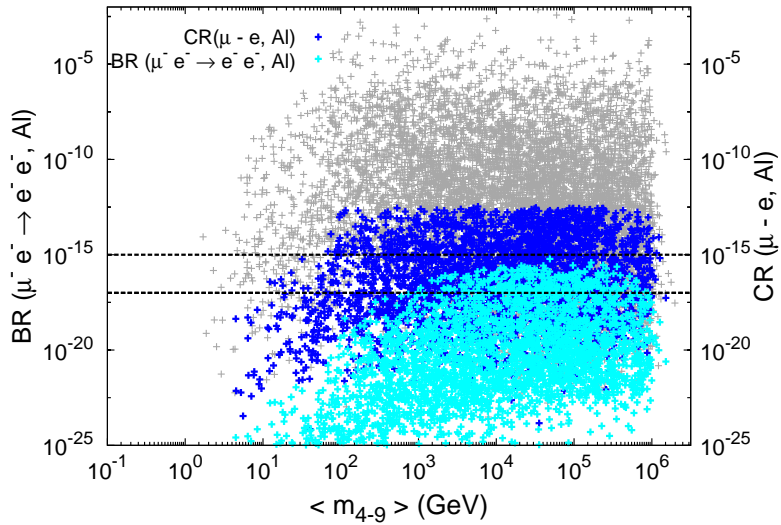,
  width=75mm, angle=270}   
\end{center}
\caption{
ISS realisation: BR($\mu^- e^- \to e^- e^-$, Al) (cyan, left axis) and 
CR($\mu -e$, Al) (dark blue, right axis)  
as a function of the average value of the mass of the 
mostly sterile states, $<m_{4-9}>$.
Grey points correspond to the violation of at least
one experimental bound. 
Dashed horizontal lines denote the future sensitivity of COMET 
(Phase I and II)~\cite{Kuno:2013mha}. } 
\label{fig:ISS_CR.BR_m49}
\end{figure}

We first consider the ISS realisation in which three RH neutrinos and
three additional steriles are added to the SM content. 
In Fig.~\ref{fig:ISS_CR.BR_m49}, we display the predictions of
the ISS concerning the Coulomb enhanced BR($\mu^- e^- \to e^- e^-$, N),
comparing it with the CR($\mu -e$, N). We consider again the case of
Aluminium targets, and display the results as a function of the
average mass of the heavier (mostly sterile) states, 
\begin{equation}
<m_{4-9}> \, =\, \frac{1}{6} \sum_{i=4...9} |m_i|\,.
\end{equation}

As can be seen, both observables are well within
experimental reach for spectra containing (at least) one pseudo-Dirac
pair even below the EW scale (concerning the coherent $\mu-e$
conversion) and above the TeV scale (for the muonic atom decay rate).
Due to the contributions of the six sterile states, one finds 
predictions for both observables 
as large as those arising in the case of the
simple ``3+1'' toy model; 
 in particular, BR($\mu^- e^- \to e^- e^-$, Al)
can reach values above $10^{-16}$, thus within reach of
COMET.

Remarkably, and despite the intrinsic constraints on its Yukawa couplings,
this (3,3) ISS realisation 
offers the following possibilities:
(i) a signal of CR($\mu -e$, Al) observable at COMET's
Phase I; (ii) BR($\mu^- e^- \to e^- e^-$, N) within the sensitivity
of COMET's Phase II even for an Aluminium target.
The ISS mass regions leading to the above signals 
can be further probed via complementarity
studies, if one again considers cLFV from the $\mu - \tau$ sector, and its
potential observation at a future collider. 
As was already the case for the simple  ``3+1'' toy model, the (3,3)
ISS regions 
with a  BR($\mu^- e^- \to e^- e^-$, Al)~$\gtrsim 10^{-17}$ would also
induce BR($Z \to \mu \tau $) within reach of the FCC-ee~\cite{Abada:2014cca}.

\begin{figure}
\hspace*{-11mm}
\begin{tabular}{cc}
\epsfig{file=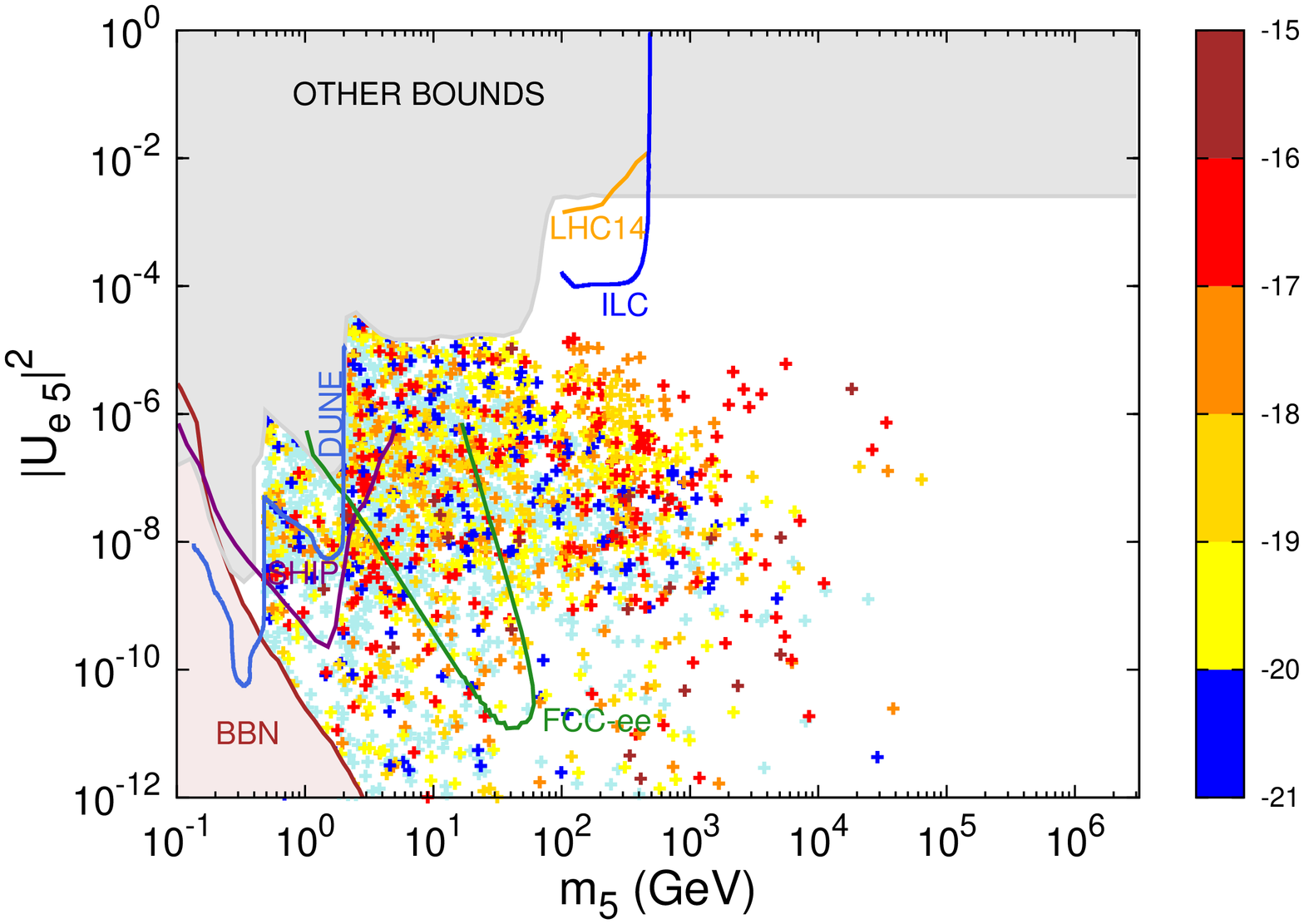,
  width=63mm, angle=270}  
\hspace*{-3mm}&\hspace*{-3mm}
\epsfig{file=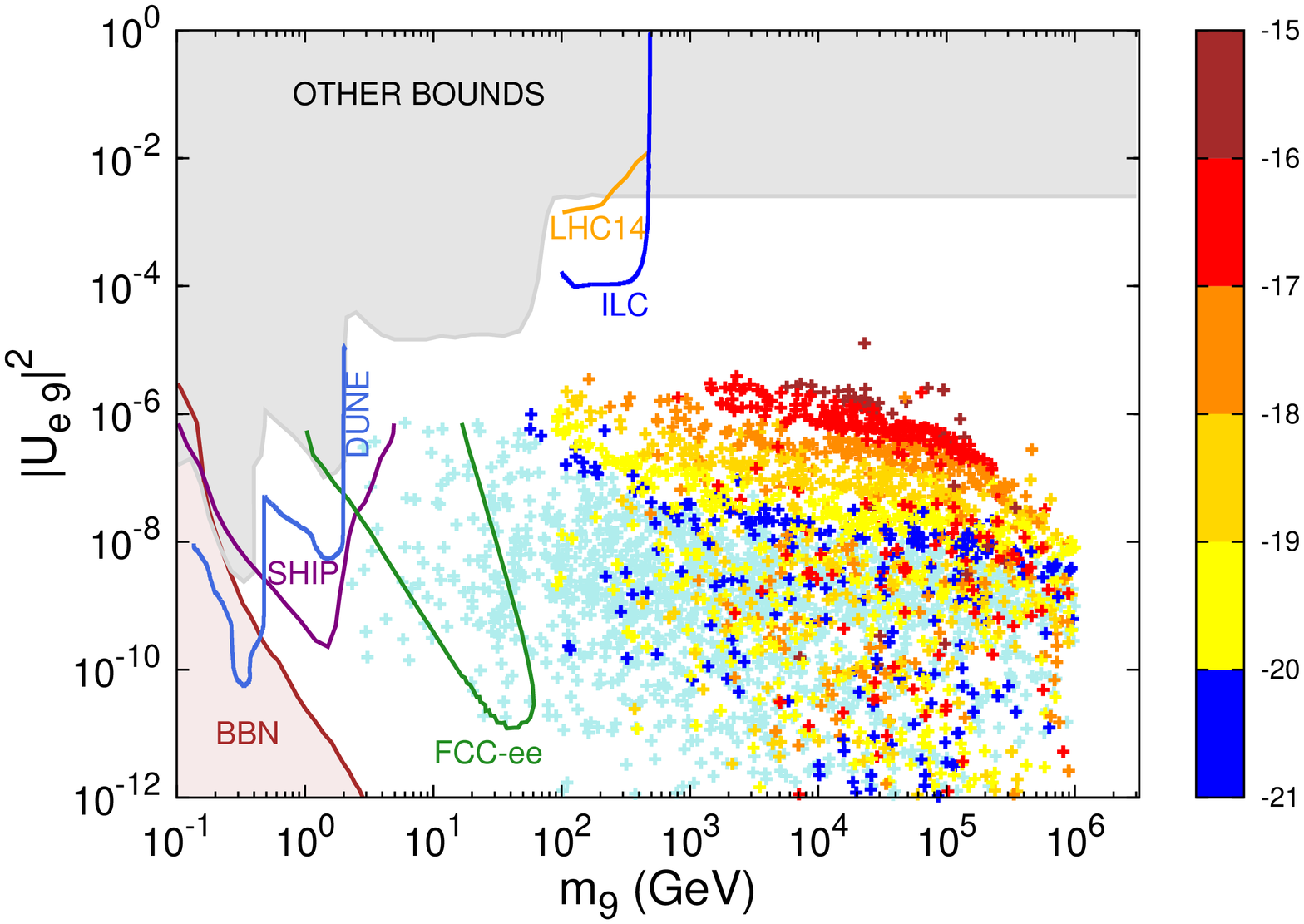,
  width=63mm, angle=270}  
\\
\epsfig{file=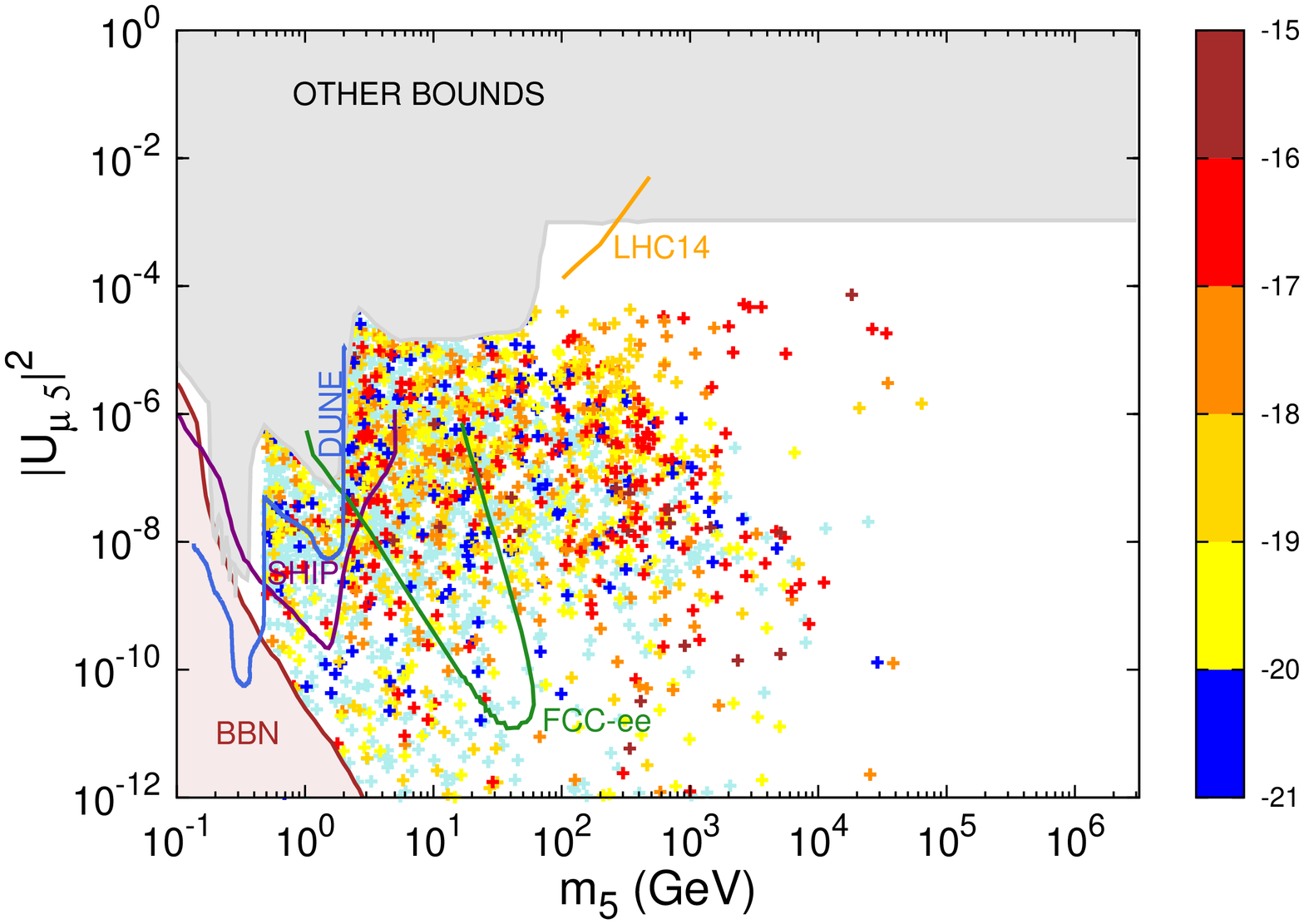,
  width=63mm, angle=270}  
\hspace*{-3mm}&\hspace*{-3mm}
\epsfig{file=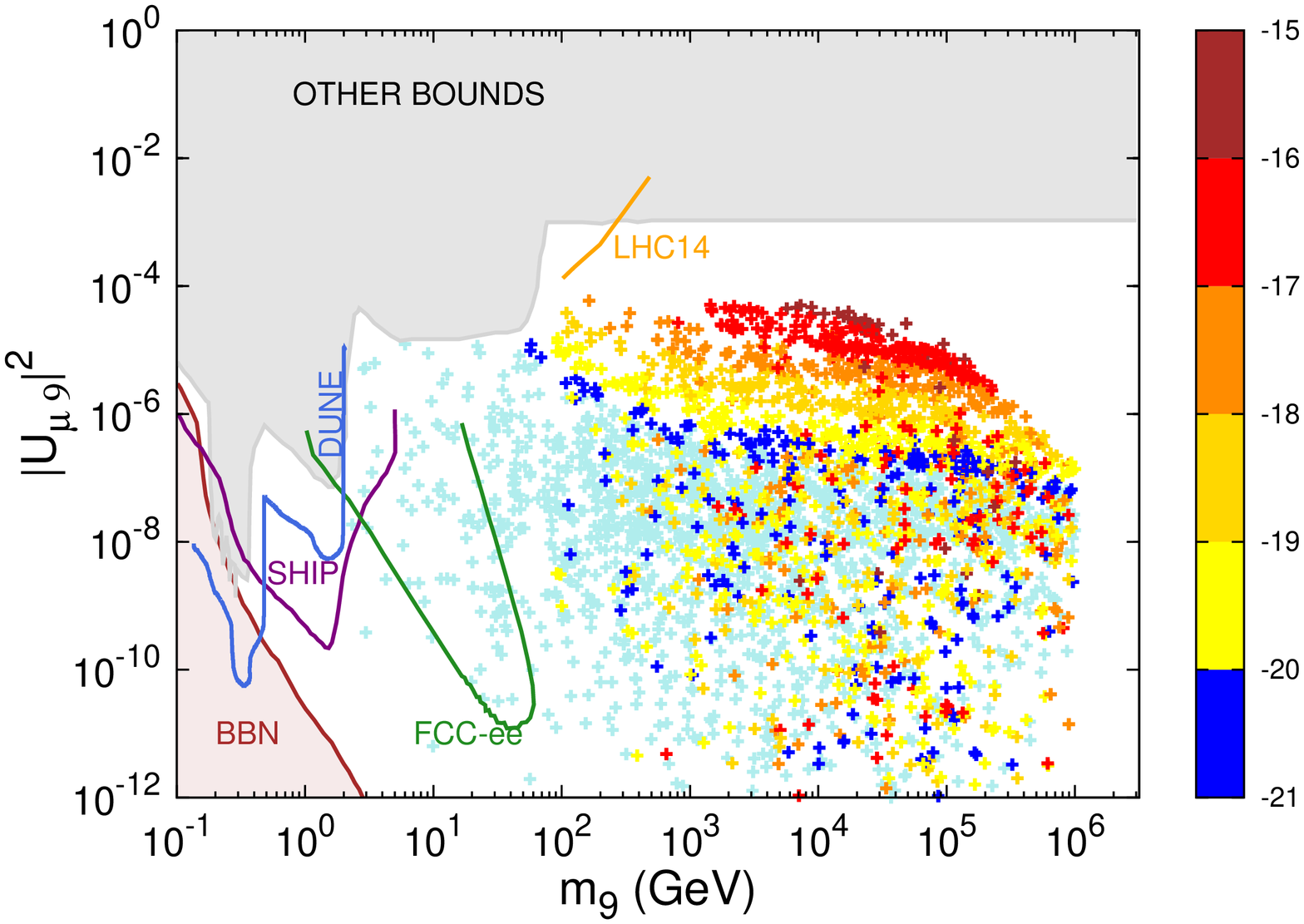,
  width=63mm, angle=270}  
\\\epsfig{file=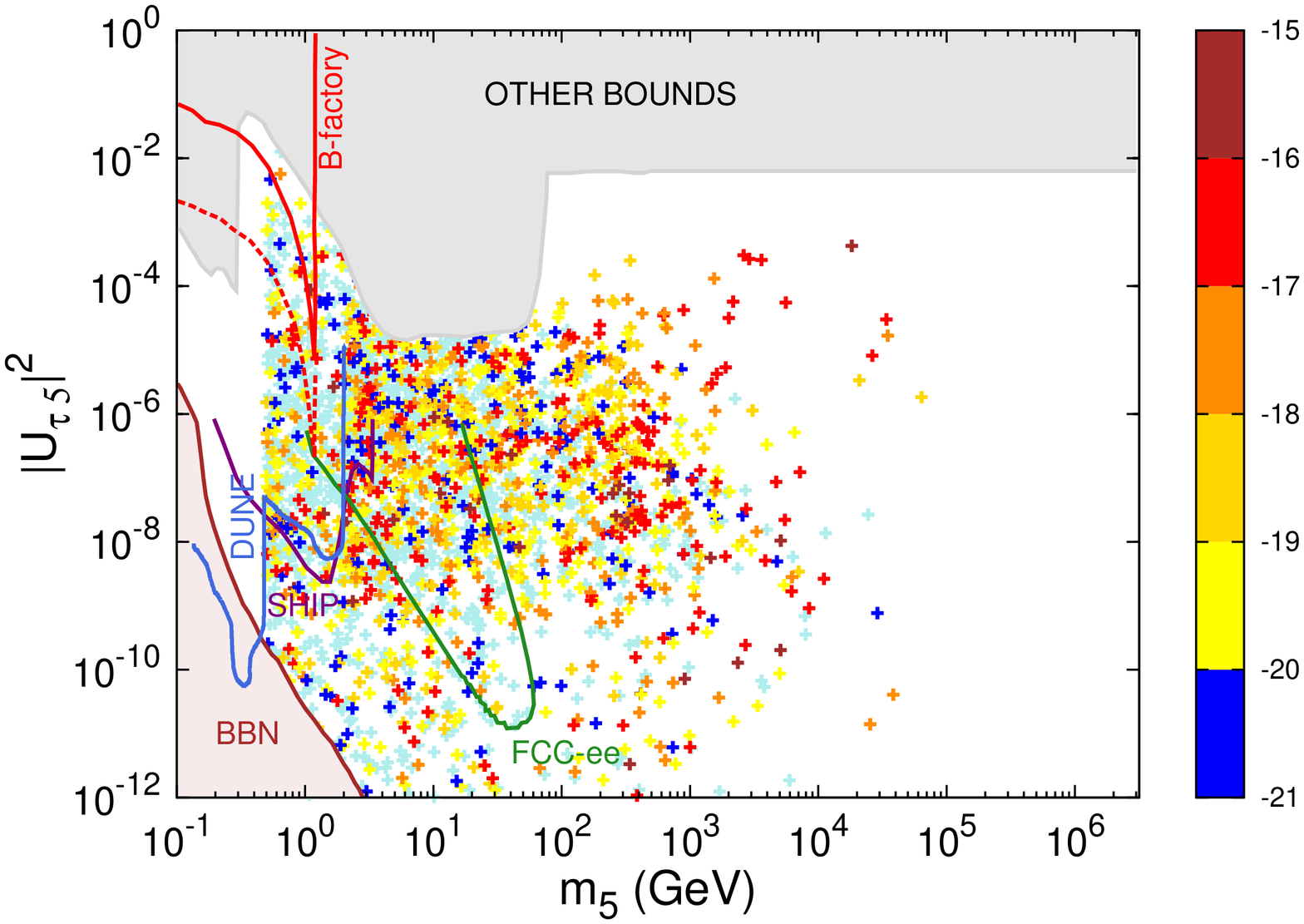,
  width=63mm, angle=270}  
\hspace*{-3mm}&\hspace*{-3mm}
\epsfig{file=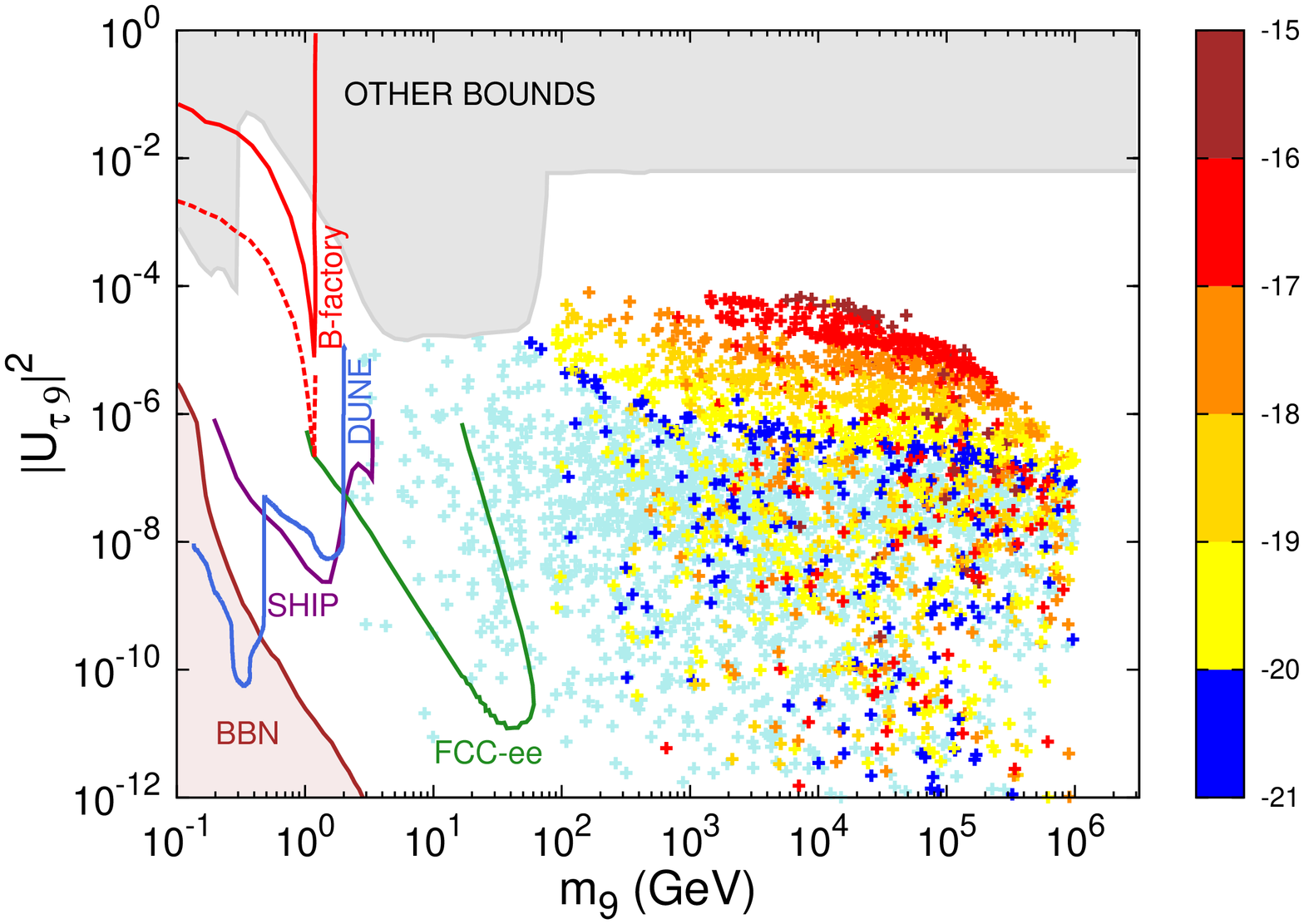,
  width=63mm, angle=270}  
\end{tabular}
\caption{ISS realisation: logarithm of BR($\mu^- e^- \to e^- e^-$, Al),
  displayed on the $(|U_{\ell i}|^2, m_i)$ parameter
  space. On the left (right) column, $i=5 \,(9)$; from top to bottom,
  $\ell=e, \, \mu$ and $\tau$.
  Line and colour code as in Fig.~\ref{fig:EFF_U2m4.palette}, with
  additional lines for the sensitivity of a future ILC in
  $(|U_{e i}|^2, m_i)$ panels (solid blue) and for a B-factory in
  $(|U_{\tau i}|^2, m_i)$ 
  (conservative limit - solid red and optimistic expectations - dashed
  red).} 
\label{fig:ISS_UmuM59}
\end{figure}

A general overview of the (3,3) ISS realisation here studied is given
in Fig.~\ref{fig:ISS_UmuM59}, in which we display the logarithm of the
BR($\mu^- e^- \to e^- e^-$, Al) in the 
$(|U_{\ell i}|^2, m_i)$ parameter space of two states: one belonging to the
lighest pseudo-Dirac pair ($\nu_5$), and another to the heaviest pair, $\nu_9$. 
In addition to the expected sensitivity reach of the experiments
already mentioned in Section~\ref{sec:res:3+1:mue2e}, 
two additional curves are depicted in the panels of
Fig.~\ref{fig:ISS_UmuM59}: the future linear collider (ILC) expected
sensitivity (solid blue) in $(|U_{e i}|^2, m_i)$ panels, which has
been obtained assuming a $\sqrt{s} = 500$ GeV and luminosity of 
$500$~fb$^{-1}$~\cite{Deppisch:2015qwa,Banerjee:2015gca}; 
the conservative and optimistic (solid red and dashed red
respectively) projected limits at 90\% C.L.  in $(|U_{\tau i}|^2,
m_i)$ panels, from semileptonic tau decays at a future 
B-factory~\cite{Kobach:2014hea}. 

As can be seen on the different panels of the left column, the states
of the lightest pseudo-Dirac pair ($\nu_{4,5}$), belonging to a sterile 
spectrum responsible for sizable BR($\mu^- e^- \to e^- e^-$, Al), 
are well within reach of the different future facilities which
are directly looking for these heavy sterile states. 
On the right column, one finds the summary 
of the heavier state's properties
(mass and couplings - displaying $U^2_{\ell 9}$ on the different
panels); this confirms the
information of Fig.~\ref{fig:ISS_CR.BR_m49} - one finds diagonal bands
corresponding to increasing values of the BR($\mu^- e^- \to e^- e^-$, Al),
for larger values of the masses and couplings. In this case, only a
small subset of the heavier states could be directly probed at FCC-ee
through $Z\to \nu_\ell \nu_9$ decays. However, and as previously
stressed, many of  
these states can be indirectly probed at the FCC-ee through
the LFV $Z \to \mu \tau$ decays. 
For the next-to heaviest states,
$\nu_{6,7}$, one encounters an intermediate scenario, with a
significant subset of the points within reach of future facilities.

\begin{figure}
\hspace*{-11mm}
\epsfig{file=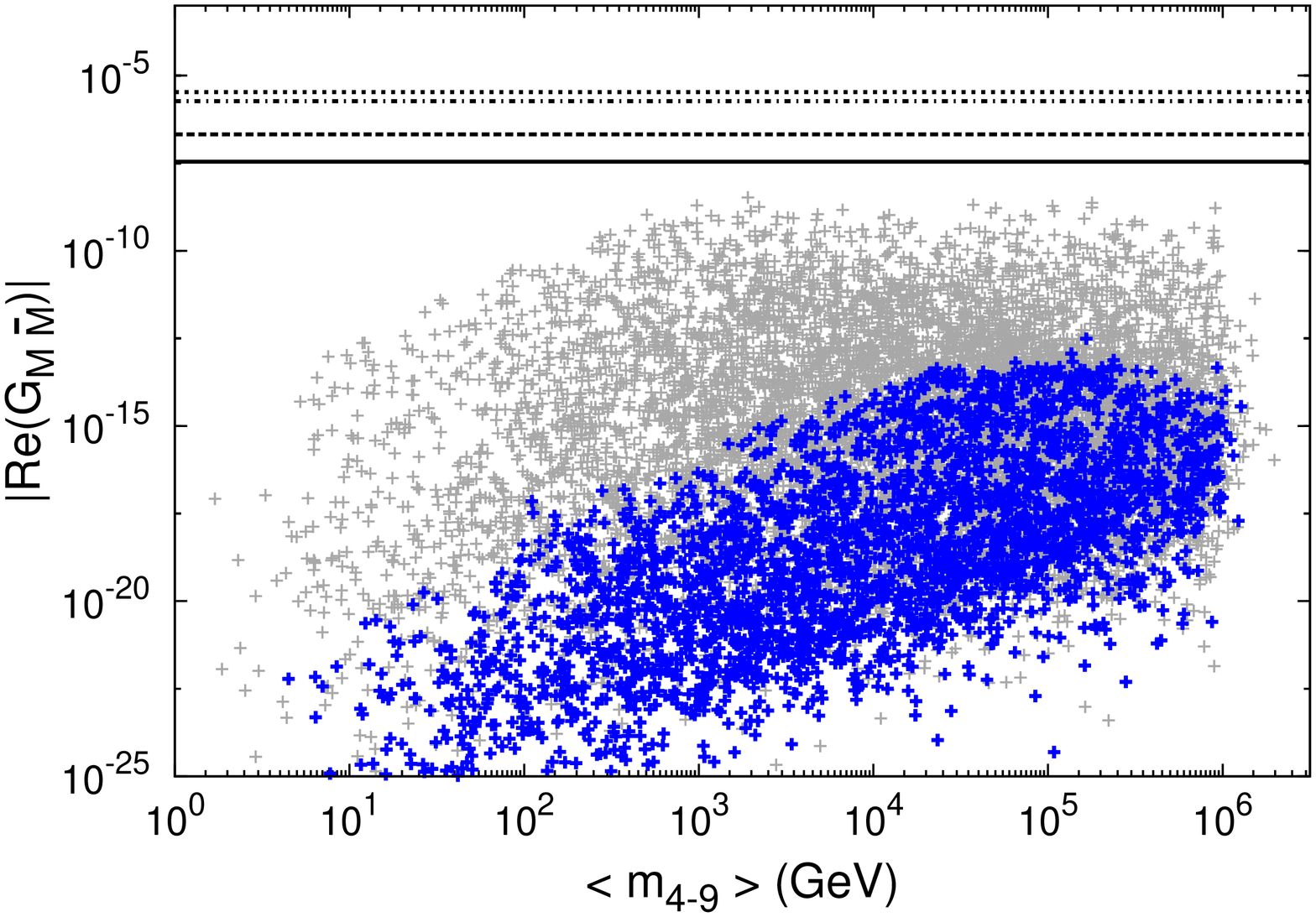,
  width=63mm,  angle=270}  
\epsfig{file=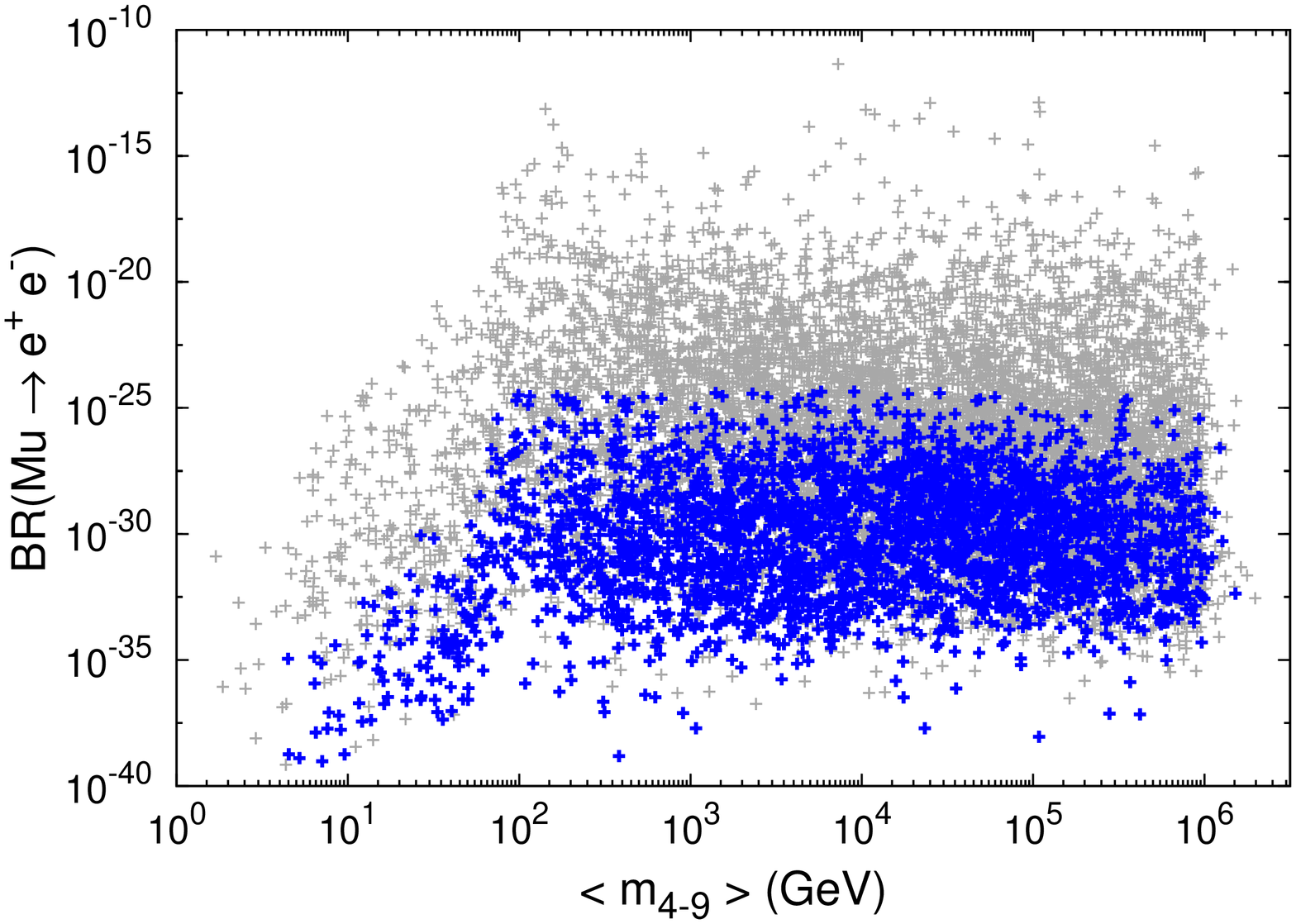,
  width=63mm,  angle=270}  
\caption{ISS realisation:  Mu - $\rm \overline{Mu}$ conversion
effective coupling (left) and  
branching ratio BR$(\text{M} \rightarrow e^+ e^-)$ (right), 
as a function of $<m_{4-9}>$. Line and colour code as in 
Fig.~\ref{fig:EFF_muonium_m4}. }
\label{fig:ISS_muonium_m49}
\end{figure}

\bigskip
Displayed in Fig.~\ref{fig:ISS_muonium_m49}, the prospects for the
observation of the $|\Delta L_{e,\mu}| =2$ Muonium conversion
(left) and of the cLFV Muonium decay (right) 
are similar to those identified in the simple ``3+1'' model; the
phenomenologically allowed regions of the parameter space would
typically lead to 
$\left |\text{Re}\left( G_{\rm M\overline{M}} \right) \right| 
\lesssim 10^{-13}$ and to  BR$(\text{Mu} \to e^+ e^-) \sim 10^{-25}$. 

\subsubsection{The $\nu$MSM}\label{sec:res:others:nuMSM}
\begin{figure}
\begin{center}
\epsfig{file=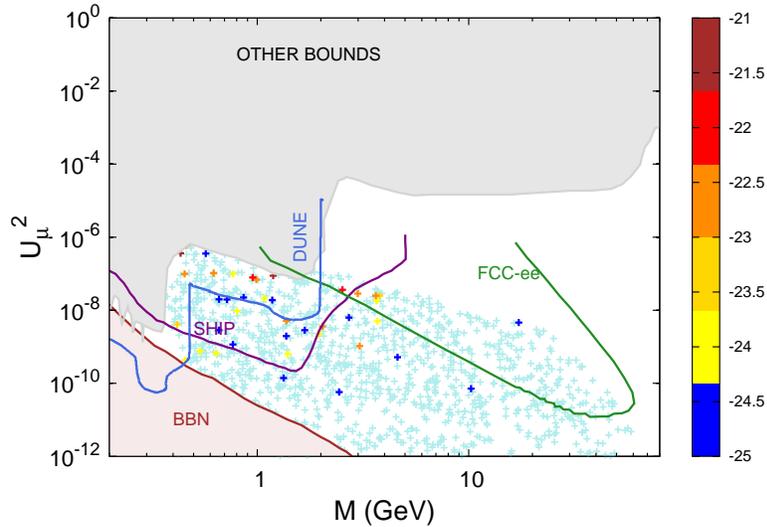,
  width=75mm,  angle=270}  
\caption{$\nu$MSM ($U^2_\mu, M$) parameter space: 
maximal values of the logarithm of the 
BR($\mu^- e^- \to e^- e^-$, N) for Uranium nuclei. 
Line and colour code as in Fig.~\ref{fig:EFF_U2m4.palette}.} 
\label{fig:numsms_all}
\end{center}
\end{figure}

Figure~\ref{fig:numsms_all} summarises the prospects of the 
$\nu$MSM concerning the decay of muonic atoms to $e^- e^-$
pairs. The results are displayed in the $(U_{\mu}^2, M)$
parameter space, generated by
the mass of the lightest (mostly sterile) state, $M \equiv m_4$, and
the mixings of the additional sterile states to the muon-neutrino,
\begin{equation}
U^2_{\mu}\, =\, \sum_{i=4}^6 \sin^2 \theta_{\mu i}\,.
\end{equation} 
At most, and in the case of Aluminium targets, 
one can expect 
 BR($\mu^- e^- \to e^- e^-$, Al)~$\sim
10^{-25}$ for a small subset of the investigated points; for heavier
targets, such as Uranium, corresponding to the results displayed in
Fig.~\ref{fig:numsms_all}, values of BR($\mu^- e^- \to e^- e^-$, U)~$\sim
10^{-22}$ can be reached.   
Although not displayed here, the rates for the coherent $\mu
-e$ conversion were also found to be very small. 
This implies that the 
$\nu$MSM's parameter space is beyond experimental
sensitivity concerning this class of nucleus-assisted cLFV observables. 
In fact, the smallness of the contributions to cLFV observables (among them those here studied) is characteristic of the $\nu$MSM given the smallness of the associated neutrino Yukawa couplings.

Concerning Mu-$\overline{\text{Mu}}$ oscillations and cLFV 
Muonium decay, the $\nu$MSM's
predictions are also extremely tiny:  
$\left |\text{Re}\left( G_{\rm M\overline{M}} \right) \right| \lesssim 10^{-26}$ and BR$(\text{Mu}
\rightarrow e^+ e^-) \sim 10^{-35}$. 

\section{Conclusions}\label{sec:concs}
In this work we have investigated the impact of sterile
fermions on cLFV observables which occur in the
presence of ``muonic atoms'': 
coherent $\mu -e$ conversion, the (Coulomb enhanced) decay of muonic
atoms into $e^- e^-$ pairs as well as  Mu-$\overline{\text{Mu}}$
oscillations and cLFV Muonium decays. 

We have considered extensions of the SM which add one or more sterile
neutrinos to its particle content: these include 
a simple ``3+1'' toy model or well motivated frameworks for
the generation of neutrino masses, for instance low-scale seesaw
mechanisms like the ISS and the $\nu$MSM. Due to their non-negligible
mixing with the lighter (mostly active) neutrinos - which induces a
departure from unitarity of the $\tilde U_\text{PMNS}$ -,
the sterile states can have a significant phenomenological impact. In
particular, they can provide non-negligible contributions to a number
of cLFV observables which are being actively searched for. In turn,
the new states are subject to a 
vast array of observational, experimental and theoretical constraints,
which must be taken into account. 

Our analysis confirmed that minimal extensions of the SM, such as the
``3+1'' toy model, can account for sizeable contributions to CR($\mu
-e$, N) and also to muon three body decays, $\mu
\rightarrow e e e$ (which is in general correlated with 
$\mu-e$ conversion). In fact, important regions of the parameter space
can be probed in near
future experiments such as Mu2e, DeeMe and COMET, while other  
regions are actually already excluded by current experimental limits on
CR($\mu -e$, Au) and BR($\mu \rightarrow e e e$). As expected, a
similar scenario arises in the context of the ISS, which  
also predicts that a signal of CR($\mu -e$, Al)
could be potentially seen at COMET in its first phase. 

The decay of a muonic atom into a pair of electrons has proven to be 
another powerful probe for cLFV,  
particularly sensitive to the presence of sterile neutrinos.
While similar to $\mu \rightarrow e e e$ from the point of view of the
elementary flavour violating transitions, this process has its rate
strongly enhanced by Coulomb interactions between the charged leptons
and the electromagnetic field of the nucleus (augmenting with increasing  
$Z$). The experimental relevance of this observable is manifest even
for the simple ``3+1'' toy model: sterile neutrinos with masses
 $m_4 \gsim$~800 GeV,
lead to BR($\mu^- e^- \to e^- e^-$, Al) within the
reach of COMET, and the contributions would be further enhanced 
for heavier atoms, such as Lead or Uranium, 
thus improving the experimental potential.  

The ISS also offers an interesting scenario concerning this
observable. Not only the added
contributions of the extra six sterile neutrinos could give rise to 
large BR($\mu^- e^- \to e^- e^-$, Al) - as large as
 $10^{-16}$ -
but for an ISS spectrum containing at least one heavy pseudo-Dirac
pair (with a mass above the TeV scale), 
one should have
signals for both BR($\mu^- e^- \to e^- e^-$, Al) and CR($\mu -e$, Al),
observable at COMET's Phase II. This ISS mass region can be
(complementary) probed 
via the search for cLFV $Z \rightarrow \mu \tau$ decays, which are
expected to be within reach of the future FCC-ee. Moreover, the
lighter pseudo-Dirac pairs, belonging to a spectrum responsible for
the large values of the cLFV observables, can also be directly searched
for at facilities such as SHiP, DUNE and FCC-ee. 

Sterile neutrinos could also contribute to
Mu-$\overline{\text{Mu}}$ conversion, and to its cLFV decay 
$\text{Mu} \to e^+ e^-$. 
Although one could in principle have values of the
effective coupling constant which would not be far from the most recent
bounds, other cLFV observables, such as CR($\mu -e$, Au) and BR($\mu
\rightarrow e e e$) preclude this possibility, and one is led to 
maximal values $\left
|\text{Re}\left(G_{\rm M\overline{M}} \right) \right| \sim 10^{-13}$ in the ``3+1'' toy model and in the
ISS, far from current experimental sensitivity.    
For completeness, 
we have also provided the expectations of these SM
extensions regarding the Muonium decay. Although the experimental roadmap is not clear at present, 
it is possible that the cLFV Muonium decay will also be included in COMET's physics programme.

Concerning the 
$\nu$MSM, it was found that the predicted 
rates for the coherent $\mu -e$ conversion
in Aluminium remain in general short of the expected future
experimental sensitivity. 
Likewise, the expected contributions to the  
BR($\mu^- e^- \to e^- e^-$, N) lie beyond experimental reach. 

To summarise, our study has shown that minimal extensions of the SM with
sterile neutrinos allow to explain future NP signals at
high-intensity facilities dedicated to search for nuclei-assisted 
cLFV observables.
In particular, that will be the case of 
CR($\mu -e$, N), BR($\mu^- e^- \to e^- e^-$, N), the latter proving to
be a very interesting probe of these SM extensions, in particular
concerning the inverse seesaw model. 
On the contrary, the potential
observation of Mu-$\overline{\text{Mu}}$ conversion or cLFV decay 
in the near future would disfavour
sterile states (as those here considered) as the unique 
source of lepton flavour violation responsible for the latter cLFV
observables.

\section*{Acknowledgements}
We are indebted  to J. Orloff for many fruitful discussions; 
we are also grateful to Y. Kuno for important remarks and contributions. 
A.M.T. is grateful to the Organisers of the ``17$^\text{th}$
COMET International Collaboration Meeting''.   
We acknowledge support from the European Union FP7 ITN
INVISIBLES (Marie Curie Actions, PITN-\-GA-\-2011-\-289442). 
This work was done in the framework of a ``D\'efi InPhyNiTi'' 
project (N2P2M-SF).

\appendix
\section{Appendix: Form factors for the cLFV decays}
\label{app:A}

In this Appendix we provide the complete analytical expressions for 
the form factors and the loop functions entering the amplitudes
of the cLFV processes described in Section~\ref{sec:cLFV:muonic}.

\subsection{Muon-electron conversion}

The form factors corresponding to the dipole, 
penguin (photon and $Z$) and box
diagrams of Fig.~\ref{Fig:muediagrams} presented in Sec.~\ref{sec:cLFV:muonic} are given
by~\cite{Alonso:2012ji,Ilakovac:1994kj,Ma:1979px,Gronau:1984ct}:  
\begin{eqnarray}
\label{eq:FF}
G^{\mu e }_\gamma &=& \sum_{j=1}^{3 + n_S} {\bf U}_{ej}{\bf U}^*_{\mu
  j} G_\gamma(x_j)\,,  \nonumber\label{Ggammamue} \\ 
F^{\mu e }_\gamma &=& \sum_{j=1}^{3 + n_S} {\bf U}_{ej}{\bf U}^*_{\mu
  j} F_\gamma(x_j)\,, \nonumber \label{Fmue}\\ 
F^{\mu e }_Z &=& \sum_{j,k=1}^{3 + n_S} {\bf U}_{ej}{\bf U}^*_{\mu k}
\left(\delta_{jk} F_Z(x_j) + {\bf C}_{jk} G_Z(x_j,x_k) + {\bf
  C}^*_{jk} H_Z(x_j,x_k)   \right)\,, \nonumber \\ 
F^{\mu e uu}_{\rm Box}&=&\sum_{j=1}^{3 + n_S}\sum_{d_\alpha=d,s,b}
{\bf U}_{ej}{\bf U}^*_{\mu j} V_{u d_\alpha} V^*_{u d_\alpha} F_{\rm
  Box}(x_j,x_{d_\alpha})\,,   \nonumber \\
F^{\mu e dd}_{\rm Box}&=&  \sum_{j=1}^{3 + n_S}\sum_{u_\alpha=u,c,t}
{\bf U}_{ej}{\bf U}^*_{\mu j} V_{d  u_\alpha } V^*_{d u_\alpha} F_{\rm
  XBox}(x_j,x_{u_\alpha})\,,   \nonumber \\  
F^{\mu eee}_{\rm Box}&=&  \sum_{j,k=1}^{3 + n_S}{\bf U}_{e j} {\bf
  U}^*_{\mu k}\left({\bf U}_{e j}{\bf U}^*_{ek}G_{\rm
  Box}(x_j,x_k)-2\,{\bf U}^*_{e j}{\bf U}_{e k}F_{\rm
  XBox}(x_j,x_k)\right)\,, \label{Fmueee}  
\end{eqnarray}
where we have introduced the dimensionless ratio of masses, 
$x_i = \frac{m^2_{\nu_i}}{m_W^2}$, $V_{q q^\prime}$ 
is the quark CKM matrix and 
${\bf C}$ has been defined in Eq.~(\ref{eq:Cmatrix:def}).
Notice that the above expressions appear in the definition of the 
$\tilde{F}_{q}^{\mu e}$  form factor (see Eq.~(\ref{eq:tildeFqmue}));
moreover, they contribute to the BR($\mu^- e^- \to e^- e^-$, N), as
seen from Eq.~(\ref{eq:br-4fermi}), to the  BR$(\mu \to
eee)$ in Eq.~(\ref{eq:mueee}), as well as to 
the amplitude of the Muonium decay
rate, cf. Eq.~(\ref{eq:amplMudecay}). 
In the limit of light masses ($x\ll 1$), 
the form factors assume the following asymptotic behaviour:
\begin{align}
F^{\mu e }_\gamma &  \xrightarrow[x\ll 1]{}  \sum_{j=1}^{3 + n_S}{\bf
  U}_{e j}{\bf U}^*_{\mu j} \left[-x_{j}\right] \, ; &	  
 G^{\mu e }_\gamma &  \xrightarrow[x\ll 1]{} \sum_{j=1}^{3 + n_S} {\bf
   U}_{e j}{\bf U}^*_{\mu j} \left[ \frac{x_j}{4} \right] \, ; \nonumber
 \\ 
F^{\mu e }_Z & \xrightarrow[x\ll 1]{} \sum_{j=1}^{3 + n_S} {\bf U}_{e
  j}{\bf U}^*_{\mu j} \left[ x_{j} \left( \frac{-5}{2}- \ln x_{j}
  \right) \right] \, ; & 
F^{\mu eee}_{\rm  Box} & \xrightarrow[x\ll 1]{}  \sum_{j=1}^{3 + n_S}
{\bf U}_{e j} {\bf U}^*_{\mu j}\left[2~x_{j} \left(1+\ln x_{j}
  \right) \right] \, .   
\end{align}
Finally, the computation of the different amplitudes calls upon the
following loop
functions~\cite{Alonso:2012ji,Ilakovac:1994kj,Ma:1979px,Gronau:1984ct}
entering the form factors of Eq.~(\ref{Fmueee}):
\begin{align}
F_Z(x)&= -\frac{5x}{2(1-x)}-\frac{5x^2}{2(1-x)^2}\ln x \, , \nonumber
\\  
G_Z(x,y)&= -\frac{1}{2(x-y)}\left[	\frac{x^2(1-y)}{1-x}\ln x -
  \frac{y^2(1-x)}{1-y}\ln y	\right]\, ,  \nonumber \\ 
H_Z(x,y)&=  \frac{\sqrt{xy}}{4(x-y)}\left[	\frac{x^2-4x}{1-x}\ln
  x - \frac{y^2-4y}{1-y}\ln y	\right] \, , \nonumber \\ 
F_\gamma(x)&= 	\frac{x(7x^2-x-12)}{12(1-x)^3} -
\frac{x^2(x^2-10x+12)}{6(1-x)^4} \ln x	\, , \nonumber \\ 
G_\gamma(x)&=    -\frac{x(2x^2+5x-1)}{4(1-x)^3} -
\frac{3x^3}{2(1-x)^4} \ln x \, ,\label{Ggamma}  \nonumber \\	
F_{\rm Box}(x, y) &= \frac{1}{x - y} \bigg\{
\left(4+\frac{x  
y}{4}\right)\left[\frac{1}{1-x}+\frac{x^2}{(1-x)^2}\ln
x\right] - 2x 
y\left[\frac{1}{1-x}+\frac{x}{(1-x)^2}\ln
x\right] -(x\to y)\bigg\} \,,\nonumber \\
F_{\rm XBox}(x, y) &= \frac{-1}{x - y} \bigg\{
\left(1+\frac{x
y}{4}\right)\left[\frac{1}{1-x}+\frac{x^2}{(1-x)^2}\ln
x\right] - 2x
y\left[\frac{1}{1-x}+\frac{x}{(1-x)^2}\ln
x\right] -(x\to y)\bigg\} \,. 			
 \end{align}
In the limit of light masses ($x\ll 1$) and/or degenerate propagators
($x=y$), one has 
\begin{align}
F_Z(x) &  \xrightarrow[x\ll 1]{}    -\frac{5x}{2} \,,\nonumber\\
G_Z(x,x ) &= {} -\left[x (-1 + x - 2 \ln x)/(2 (x -1)) \right]]\, , \,\,
G_Z(x,x)  \xrightarrow[x\ll 1]{}  -\frac{1}{2} x \ln x \,, \nonumber\\
H_Z(x,x ) & = {} - \left[ \sqrt{x^2} (4 - 5x + x^2 + (4 - 2x + x^2)\ln
  x)/(4(x - 1)^2) \right] \, , \nonumber\\ 
F_\gamma(x ) & \xrightarrow[x\ll 1]{}  -x \,, \nonumber\\
G_\gamma(x )  & \xrightarrow[x\ll 1]{}  \frac{x}{4}\, \nonumber\\
 F_{\rm Box}(x,x)	 &= \left[\left(-16 + 31x^2 - 16x^3 + x^4 +
   2x(-16 + 4x + 3x^2) \ln x \right)/\left(4(-1 + x)^3\right) \right]
 \, , \nonumber\\ 
F_{\rm XBox}(x,x )& = \left[ (-4 + 19 x^2 - 16 x^3 + x^4 + 2x (-4  4 x
  + 3x^2) \ln x)/(4(x - 1)^3) \right] \, .\label{limitval2} 
\end{align}

\subsection{Muonium:  Mu-$\rm \bf \overline{Mu}$ conversion and Mu decay}
The loop function, obtained from the integration of the Dirac - and
Majorana - boxes of Fig.~\ref{fig:muonium_boxes}, 
responsible for Muonium-antimuonium conversion, is~\cite{Clark:2003tv}: 
\begin{equation}
G_{\rm {Muonium}}(x_i, x_j) =  x_i x_j \left(
\frac{J(x_i)-J(x_j)}{x_i-x_j} \right)\,, 
\end{equation}
where 
\begin{equation}
J(x) = \frac{ (x^2-8x+4)}{4(1-x)^2} \ln x-\frac{3}{4}\frac{1}{(1-x)}.
\end{equation}
In the degenerate case, in which $x_i=x_j=x$, $ G_{\rm Muonium}$ is
given by
\begin{equation}
 G_{\rm Muonium}(x)  =  \frac{x^3 -11x^2
   +4x}{4(1-x)^2}-\frac{3x^3}{2(1-x)^3} \ln x\,. 
\end{equation}
Concerning the expression for the rate of the cLFV Muonium decay, the 
total amplitude (squared, summed over final spins and averaged over 
initial ones) of Eq.~(\ref{eq:Mudecay:BR}) can be cast as
\cite{Cvetic:2006yg}: 
\begin{align}
|\mathcal{M}_\text{tot}|^2\,  =\, & 
\frac{\alpha_w^4}{16 M_W^4}
\bigg\{
\left( m_e\,m_\mu^{3}+2\,{m_e}^{2}m_\mu^{2}
+{m_e}^{3}m_\mu \right) {\left| 2F_Z^{\mu e} + F_{\rm Box}^{\mu e e e}\right|}^{2}
\nonumber\\
&+ 4 \sin^2 \theta_w\left( 2\,m_e\,m_\mu^{3}
+3\,{m_e}^{2}m_\mu^{2}+3\,{m_e}^{3}m_\mu \right)\
{\rm Re} \left[(2F_Z^{\mu e}+F_{\rm Box}^{\mu e e e})
(F_\gamma^{\mu e} -F_Z^{\mu e})^*\right]\nonumber\\
&+ 12 \sin^2 \theta_w
\left( m_e\,m_\mu^{3}+ 2\,{m_e}^{2}m_\mu^{2}+{m_e}^{3}m_\mu \right)
\ {\rm Re} \left[(2F_Z^{\mu e}+F_{\rm Box}^{\mu e e e})
G_\gamma^{\mu e *}\right] \nonumber\\
 &+ 4 \sin^4\theta_w \left( 7\,m_e\,{m_\mu}^{3}+12\,{m_e}^{2}m_\mu^{2}
 +9\,{m_e}^{3}m_\mu \right) {\left|F_\gamma^{\mu e}-F_Z^{\mu
     e}\right|}^{2}\nonumber\\ 
&+ 4 \sin^4\theta_w \left( -2\,m_\mu^{4}
+12\,m_e\,m_\mu^{3}+36\,{m_e}^{2}m_\mu^{2}+18\,{m_e}^{3}m_\mu \right)
\ {\rm Re} \left[(F_\gamma^{\mu e} - F_Z^{\mu e}) G_\gamma^{\mu e*}
\right]\nonumber\\
&+ 4 \sin^4\theta_w \left( {\frac {m_\mu^{5}}{m_e}}+2\,m_\mu^{4}
+8\,m_e\,{m_\mu}^{3}+24\,{m_e}^{2}m_\mu^{2}
+9\,{m_e}^{3}m_\mu \right) \left|{G_\gamma^{\mu e}}\right|^{2}\bigg\}
\ ,
\label{eq:amplMudecay}
\end{align}
where the form factors $F^{\mu e}, F_Z^{\mu e}, G_\gamma^{\mu e},
F_{\rm Box}^{\mu e e e}$  have been given in Eq.~(\ref{eq:FF}). 

\section{Appendix: Three body muon decays $\mu \to eee$}
\label{app:B}
For completeness, we include here 
the expression for  the branching ratio 
BR$(\mu \to eee)$~\cite{Ilakovac:1994kj,Alonso:2012ji}, an observable
which was included in our main discussion (due to its constraining
r\^ole on the different parameter spaces): 
\begin{align}
\text{BR}
(\mu   \rightarrow eee)=& \frac{\alpha^4_w
}{24576\pi^3}\frac{m^4_\mu}{M^4_W}\frac{m_\mu}{\Gamma_\mu} 
\times \Bigg\{ 2 \left|\frac{1}{2}F^{\mu eee}_{\rm Box}+F^{\mu
  e}_Z-2s^2_w(F^{\mu e}_Z-F^{\mu e}_\gamma)\right|^2+4 s^4_w
\left|F^{\mu e}_Z-F^{\mu e}_\gamma\right|^2 \nonumber \\
&+ 16 s^2_w \text{Re}\left[	(F^{\mu e}_Z +\frac{1}{2}F^{\mu eee}_{\rm
Box})	G^{\mu e*}_\gamma \right]
- 48 s^4_w \text{Re}\left[(F^{\mu e}_Z-F^{\mu e}_\gamma)	G^{\mu
    e*}_\gamma \right]	\nonumber \\ 
&+32 s^4_w |G^{\mu e}_\gamma|^2\left[\ln
  \frac{m^2_\mu}{m^2_{e}} -\frac{11}{4}	\right]
\Bigg\}\,,  \label{eq:mueee} 
\end{align}
 which contains the same form factors as those entering in 
CR($\mu-e$, N), although in different combinations.

\end{document}